\documentclass[preprint,showpacs,showkeys]{revtex4-1}
\usepackage{graphicx}
\usepackage[caption=false]{subfig}
\usepackage{epsfig}
\usepackage{amssymb}
\usepackage{amsmath}
\usepackage{lipsum}
\usepackage{subfig}
\usepackage{mathrsfs}
\RequirePackage[colorlinks,citecolor=blue,urlcolor=magenta,linkcolor=red]{hyperref}
\begin{document} 
\title{Transverse Momentum Dependent Parton Distributions with self-similarity at small \textit{x} and models of proton structure function}
\author{D. K. Choudhury$^{1,2}$}
\author{Baishali Saikia$^{1}$}
\email[Corresponding author: ] {baishalipiks@gmail.com}
\affiliation{$^1$Department of Physics, Gauhati University, Guwahati 781 014, Assam, India}
\affiliation{$^2$Physics Academy of North-East, Guwahati 781 014, Assam, India}

\begin{abstract}
In this paper we make re-analysis of a self-similarity based model of the proton structure function at small \textit{x} pursued in  recent years. The additional assumption is that it should be singularity free in the entire kinematic range $0\leq \textit{x}\leq 1$. Our analysis indicates that the singularity free version of the model is valid in a more restrictive range of $Q^{2}$. We then analyse the defining Transverse Momentum Dependent Parton Distributions (TMD) occurred in the models and show that the proper generalizations and initial conditions on them not only remove the undesired singularity but also results in a QCD compatible structure function with logarithmic growth in $Q^2$. The phenomenological range of validity is then found to be much larger than the earlier versions. We also extrapolate the models to large \textit{x} in a parameter free way. 
\end{abstract}
\maketitle

\section{Introduction}
\label{intro}
\label{A}
Although renormalization group equation of quantum field theory \cite{gman} exhibits self-similarity \cite{shir}, it is not yet established rigorously in QCD, the accepted fundamental quantum field theory of strong interaction. However because of its wide applicability in other areas of physics \cite{kro,cpl,4} including condensed matter physics, its applicability in the study of structure of the proton is worth pursuing at least at phenomenological level. In the middle of 1980's, the notion of fractals has found its applicability in hadron production process \cite{5,6,7,8} when the self-similar nature of hadron multi-particle production process was suggested. Specifically in 1990, Bjorken \cite{8} highlighted the fractality of parton cascades leading to the anomalous dimension of phase space.

Relevance of these ideas in the contemporary physics of DIS has been first noted by Dremin and Levtchenko \cite{9} in early 1990's where it was shown that the saturation of hadron structure function at small \textit{x} may proceed faster if the highly packed regions of proton have fractal structures. However, it was Lastovicka\cite{Last} in 2002, who first suggested the self-similarity as a possible feature of multipartons in the proton specially in the kinematical region of small Bjorken \textit{x}, which in later years was pursued in Ref \cite{DK1,DK2,DK3,DK9,dka,dkb,DK4,DK5,DK6,DK8,dkc}. Specifically how quarks and gluons share the momentum fractions of the proton in self-similar way was studied in \cite{DK4, DK5}, large \textit{x} behavior of parton distribution functions (PDF) and double parton distribution functions (dPDF) in \citep{DK6}, and Froissart saturation in \cite{dkc}.

One of the apparent limitations of the phenomenological analysis of Ref\cite{Last} is that it has a singularity at $x_0\sim 0.019$ which is well within the kinematical range $0\leq x \leq 1$. However such singularity is not a common expectation from any physically viable model of proton structure function $F_2(x,Q^2)$.

In the present paper, we therefore make a re-analysis of the model of Ref\cite{Last}, demanding it to be singularity free in the entire $x$-range of $0\leq x \leq 1$. To that end we will use the more recently complied HERA data\cite{HERA,pdf2,z}, instead of analysis of Ref\cite{Last} where as previously reported data were used Ref\cite{H1,ZE}. In section \ref{B}, we outline the formalism, and in section \ref{C}, we discuss the results and compare the two version of the model and their limitations. A plausible way of removing the limitations of the models is suggested in section \ref{D} through a reconstruction of the defining Transverse Momentum Dependent Parton Distributions (TMDPDF). Specifically, we show that the proper generalizations and initial conditions on them not only remove the undesired singularity but also results in a QCD compatible structure function with logarithmic growth in $Q^2$. The phenomenological range of validity is then found to be much larger than the earlier versions. Section \ref{E} contains the conclusions.

\section{Formalism}
\label{B}
\subsection{Proton structure function based on self-similarity}
The self-similarity based model of the proton structure function of Ref\cite{Last} is based on transverse momentum dependent parton distribution function(TMD) $f_i(x,k_t^2)$. Here $\textit{k}_t^2$ is the parton transverse momentum squared. Choosing the magnification factors $M_1= \left(1+\frac{k_t^2}{k_0^2}\right)$ and $M_2= \left(\frac{1}{x}\right)$, it is written as \cite{Last,DK6} 
\begin{equation}
\label{E1}
\log[M^2.f_i(x,k_t^2)]= D_1.\log\frac{1}{x}.\log\left(1+\frac{k_t^2}{k_0^2}\right)+D_2.\log\frac{1}{x}+D_3.\log\left(1+\frac{k_t^2}{k_0^2}\right)+D_0^i
\end{equation}
\\
where \textit{i} denotes a quark flavor. Here $D_1,\ D_2,\ D_3$ are the three flavor independent model parameters while $D_0^i$ is the only flavor dependent normalization constant. $M^2$(=1 GeV$^2$) is introduced to make (PDF) $q_i(x,Q^2)$ as defined below (in Eqn \ref{E2}) dimensionless. The integrated quark densities then can be defined as
\begin{equation}
\label{E2}
q_i(x,Q^2) = \int_0^{Q^2}f_i(x,k_t^2)dk_t^2
\end{equation}
\\
As a result, the following analytical parametrization of a quark density is obtained by using Eqn(\ref{E2}) \cite{DK5} : \textbf{(Model 1)}
\begin{equation}
\label{E3}
q_i(x,Q^2) = e^{D_0^i}f(x,Q^2)
\end{equation}
where
\begin{equation}
\label{E4}
f(x,Q^2)= \frac{Q_0^2 \ \left( \frac{1}{x}\right) ^{D_2}}{M^2\left(1+D_3+D_1\log\left(\frac{1}{x}\right)\right)} \left(\left(\frac{1}{x}\right)^{D_1\log \left(1+\frac{Q^2}{Q_0^2}\right)} \left(1+\frac{Q^2}{Q_0^2}\right)^{D_3+1}-1 \right)
\end{equation}
\\
is flavor independent. Using Eqn(\ref{E3}) in the usual definition of the structure function $F_2(x,Q^2)$, one can get
\begin{equation}
\label{E5}
F_2(x,Q^2)=x\sum_i e_i^2 \left( q_i(x,Q^2)+ \bar{q}_i(x,Q^2)\right) 
\end{equation}
or it can be written as
\begin{equation}
\label{E6}
F_2(x,Q^2)=e^{D_0}xf(x,Q^2)
\end{equation}
\\
where 
\begin{equation}
\label{Ea}
e^{{D_0}}=\sum_{i=1}^{n_f}e^{2}_{i}\left(e^{D_0^i}+ e^{\bar{D}_0^i}\right)
\end{equation}
\\
Eqn(\ref{E5}) involves both quarks and anti-quarks. As in Ref\cite{Last} we use the same parametrization both for quarks and anti-quarks. Assuming the quark and anti-quark have equal normalization constants, we obtain for a specific flavor
\begin{equation}
\label{Eb}
e^{{D_0}}=\sum_{i=1}^{n_f}e^{2}_{i}\left(2 e^{D_0^i}\right)
\end{equation}
\\
It shows that the value of $D_0$ will increase as more and more number of flavors contribute to the structure function.\\ With $n_f=3 , 4 $ and 5 it reads explicitly as
\begin{eqnarray}
\label{Ec}
n_f=3 &:& \ \ e^{D_0}= 2 \left( \frac{4}{9}e^{{D_0}^{u}}+\frac{1}{9}e^{{D_0}^{d}}+\frac{1}{9}e^{{D_0}^{s}} \right) \\
\label{Ed}
n_f=4 &:& \ \ e^{D_0}= 2 \left(\frac{4}{9}e^{{D_0}^{u}}+\frac{1}{9}e^{{D_0}^{d}}+\frac{1}{9}e^{{D_0}^{s}}+\frac{4}{9}e^{{D_0}^{c}}\right) \\
\label{Ef}
n_f=5 &:& \ \ e^{D_0}= 2 \left(\frac{4}{9}e^{{D_0}^{u}}+\frac{1}{9}e^{{D_0}^{d}}+\frac{1}{9}e^{{D_0}^{s}}+\frac{4}{9}e^{{D_0}^{c}}+\frac{1}{9}e^{{D_0}^{b}}\right) 
\end{eqnarray}
\\
Since each term of right hand sides of Eqn(\ref{Ec}),(\ref{Ed}), and (\ref{Ef}) is positive definite, it is clear, the measured value of $D_0$ increases as $n_f$ increases. However, single determined parameter $D_0$ can not ascertain the individual contribution from various flavors.
\\ \\
From HERA data \cite{H1,ZE}, Eqn(\ref{E6}) was fitted in Ref\cite{Last} with
\begin{eqnarray}
\label{E7}
D_0 &=& 0.339\pm 0.145 \nonumber \\
D_1 &=& 0.073\pm 0.001 \nonumber \\
D_2 &=& 1.013\pm 0.01 \nonumber \\
D_3 &=& -1.287\pm 0.01 \nonumber \\
Q_0^2 &=& 0.062\pm 0.01 \ {\text G\text e\text V^2}
\end{eqnarray}

in the kinematical region,
\begin{eqnarray}
\label{E8}
& & 6.2\times10^{-7}\leq x\leq 10^{-2} \nonumber \\
& & 0.045\leq Q^2 \leq 120 \ {\text G\text e\text V^2}
\end{eqnarray}

\subsection{Singularity free structure function}
The defining equations of the model of Ref\cite{Last} (Eqn \ref{E1}-\ref{E4} above) do not ascertain the numerical values and signs of the parameters $D_j$ s. These are determined from data\cite{H1,ZE} leading to the set of Eqn(\ref{E7}) in the kinematic range (Eqn \ref{E8}). However, the phenomenological analysis has one inherent limitation: due to the  negative value of $D_3$, Eqn(\ref{E6}) develops a singularity at $x_0 \backsim 0.019$ \cite{DK4, DK5} as it satisfies the condition $1+D_3+D_1\log\frac{1}{x_0}=0$, contrary to the expectation of a physically viable form of structure function.

Redefining the model parameters $D_j$ s by $D'_j$ s (\textit{j}=1,2,3) and (PDF) $q_i(x,Q^2)$ by $q'_i(x,Q^2)$ and also structure function $F_2(x,Q^2)$ by $F'_2(x,Q^2)$ in the present model, we get the following forms of PDF and structure function as : \textbf{(Model 2)}

\begin{equation}
\label{a1}
q_i'(x,Q^2)=\frac{e^{D_0'^i}\ Q_0'^2\ \left( \frac{1}{x}\right) ^{D'_2}}{M^2 \left(1+D'_3+D'_1\log\frac{1}{x}\right)} \left(\left(\frac{1}{x}\right)^{D'_1\log\left(1+\frac{Q^2}{Q_0'^2}\right)}\left(1+\frac{Q^2}{Q_0'^2}\right)^{D'_3+1}-1\right)
\end{equation}
and
\begin{equation}
\label{a2}
F'_2(x,Q^2)=\frac{e^{D'_0}\ Q_0'^2\ \left( \frac{1}{x}\right) ^{D'_2-1}}{M^2 \left(1+D'_3+D'_1\log\frac{1}{x}\right)} \left(\left(\frac{1}{x}\right)^{D'_1\log\left(1+\frac{Q^2}{Q_0'^2}\right)}\left(1+\frac{Q^2}{Q_0'^2}\right)^{D'_3+1}-1\right)
\end{equation}
respectively.
\section{Results}
\label{C}
\subsection{Analysis of singularity free model}
To determine the model parameters $\left(D'_0 , D'_1 , D'_2 , D'_3 , Q_0'^2 \right)$  we have used the compiled HERA data \cite{HERA} instead of earlier data \cite{H1,ZE}  used in Ref\cite{Last}. The more recent HERA communication \cite{pdf2,z} do not give additional new information of structure function with the kinematical region. Following the procedure of Ref\cite{Last}, we make $\chi^2$-analysis of the data and obtained the more restrictive range of $Q^2$ and \textit{x} : $0.85\leq Q^2 \leq$ 10 GeV$^2$ and $2\times10^{-5}\leq x\leq 0.02$ respectively with the  fitted parameters given in the Table \ref{Table4}. The number of data points of $ F'_2$ is 95.

\begin{figure*}[!tbp]
 \captionsetup[subfigure]{labelformat=empty}
 \centering
  \subfloat[]{\includegraphics[width=.3\textwidth]{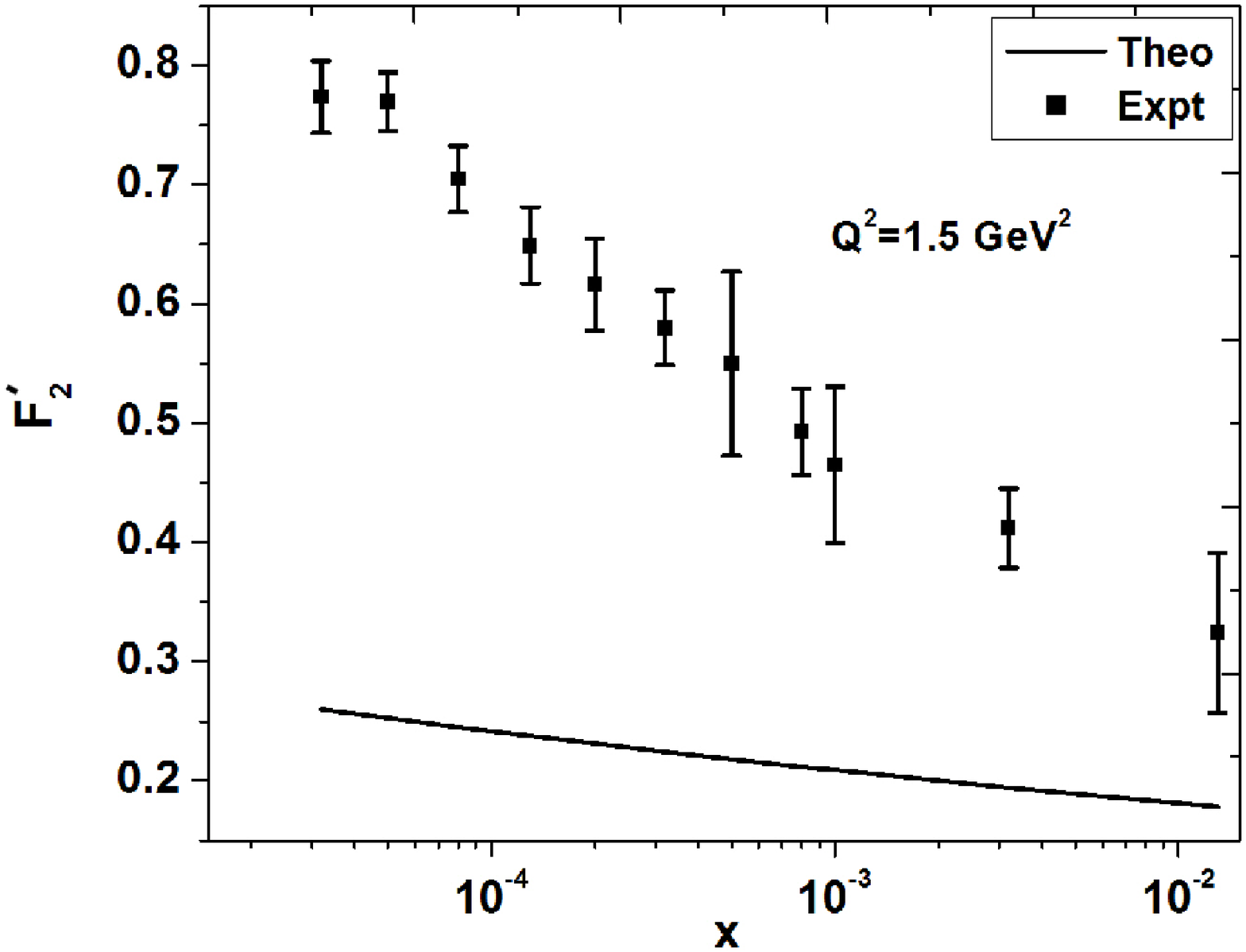}}\quad
\subfloat[]{\includegraphics[width=.3\textwidth]{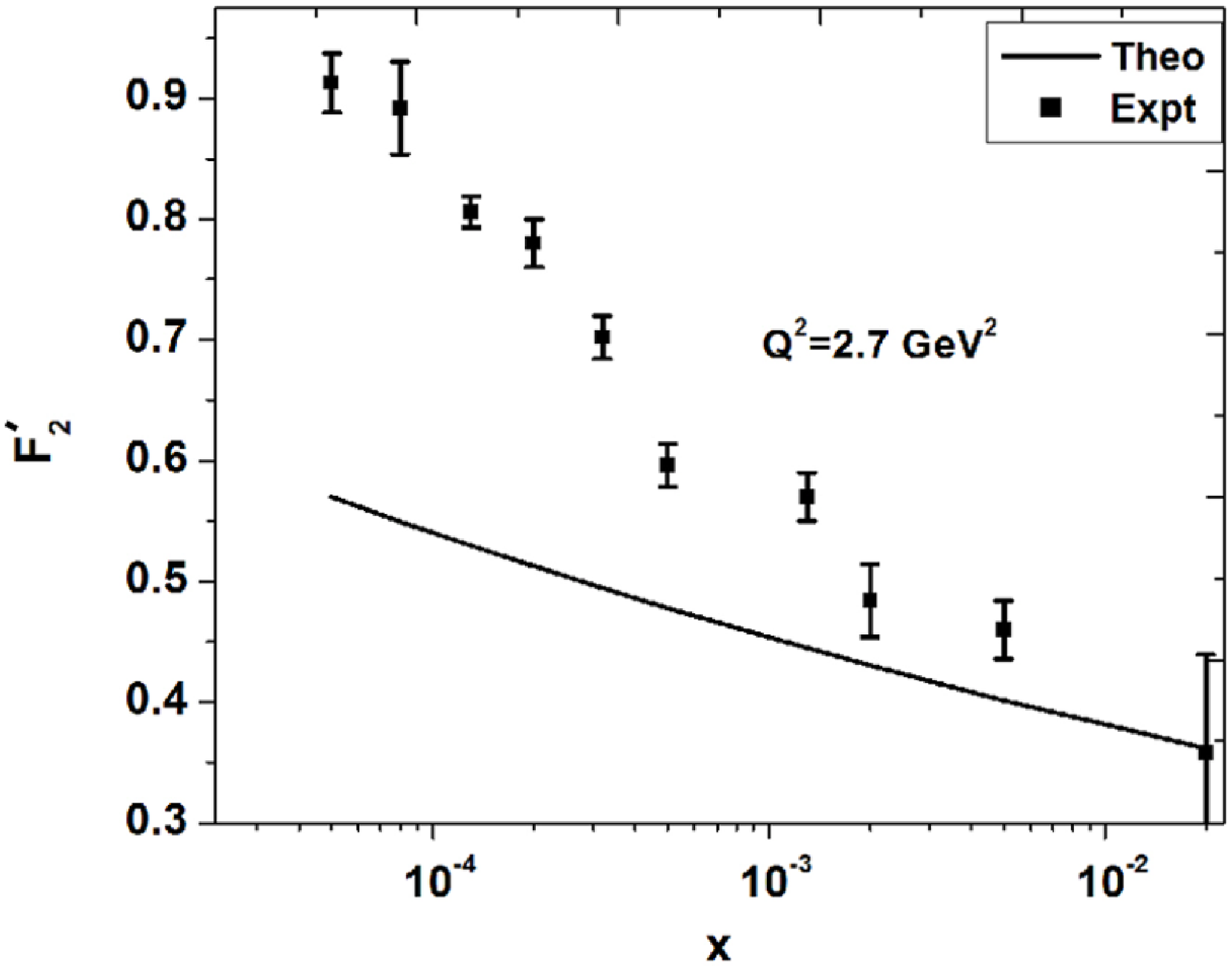}}\quad
 \subfloat[]{\includegraphics[width=.3\textwidth]{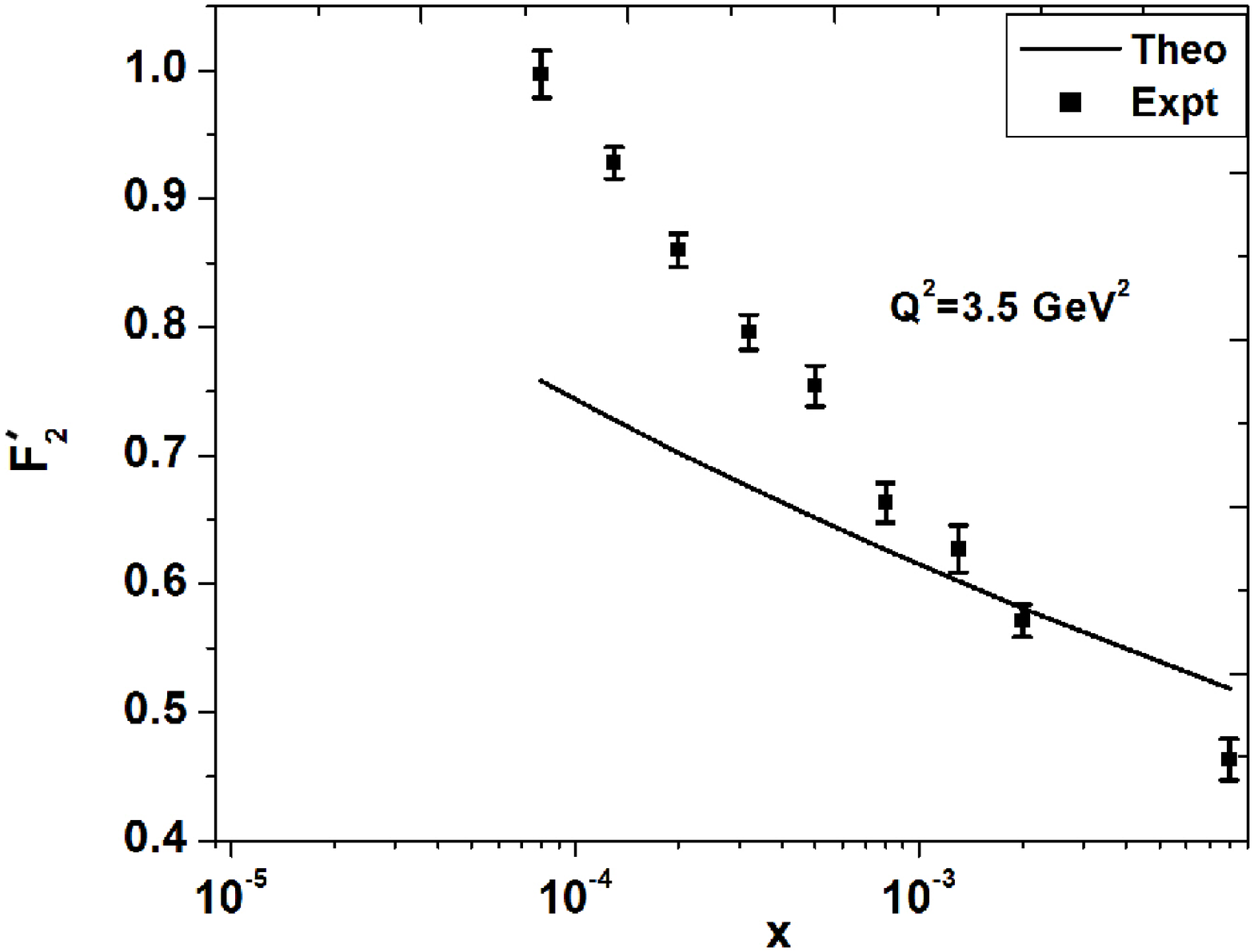}}\quad
\subfloat[]{\includegraphics[width=.3\textwidth]{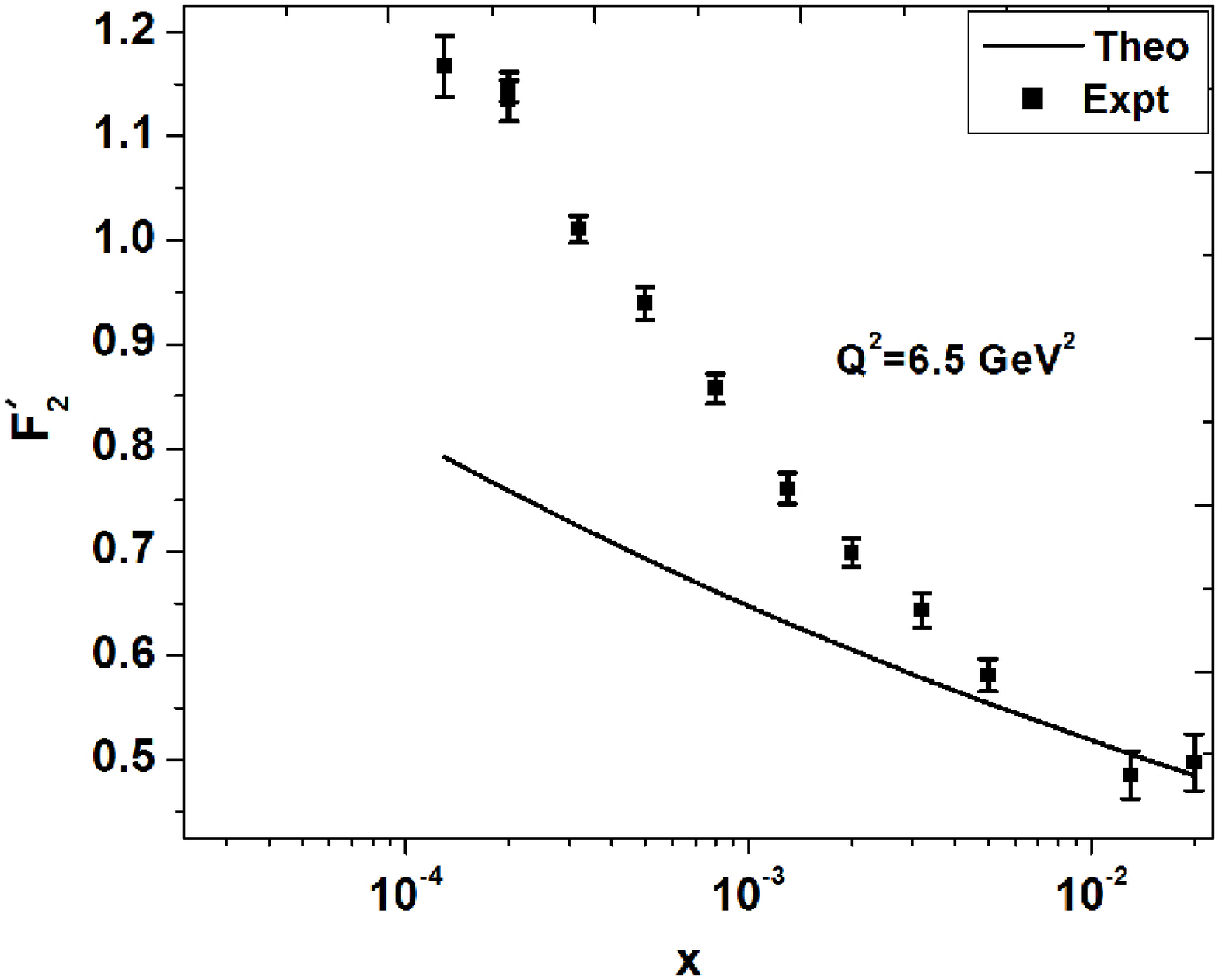}}\quad
\subfloat[]{\includegraphics[width=.3\textwidth]{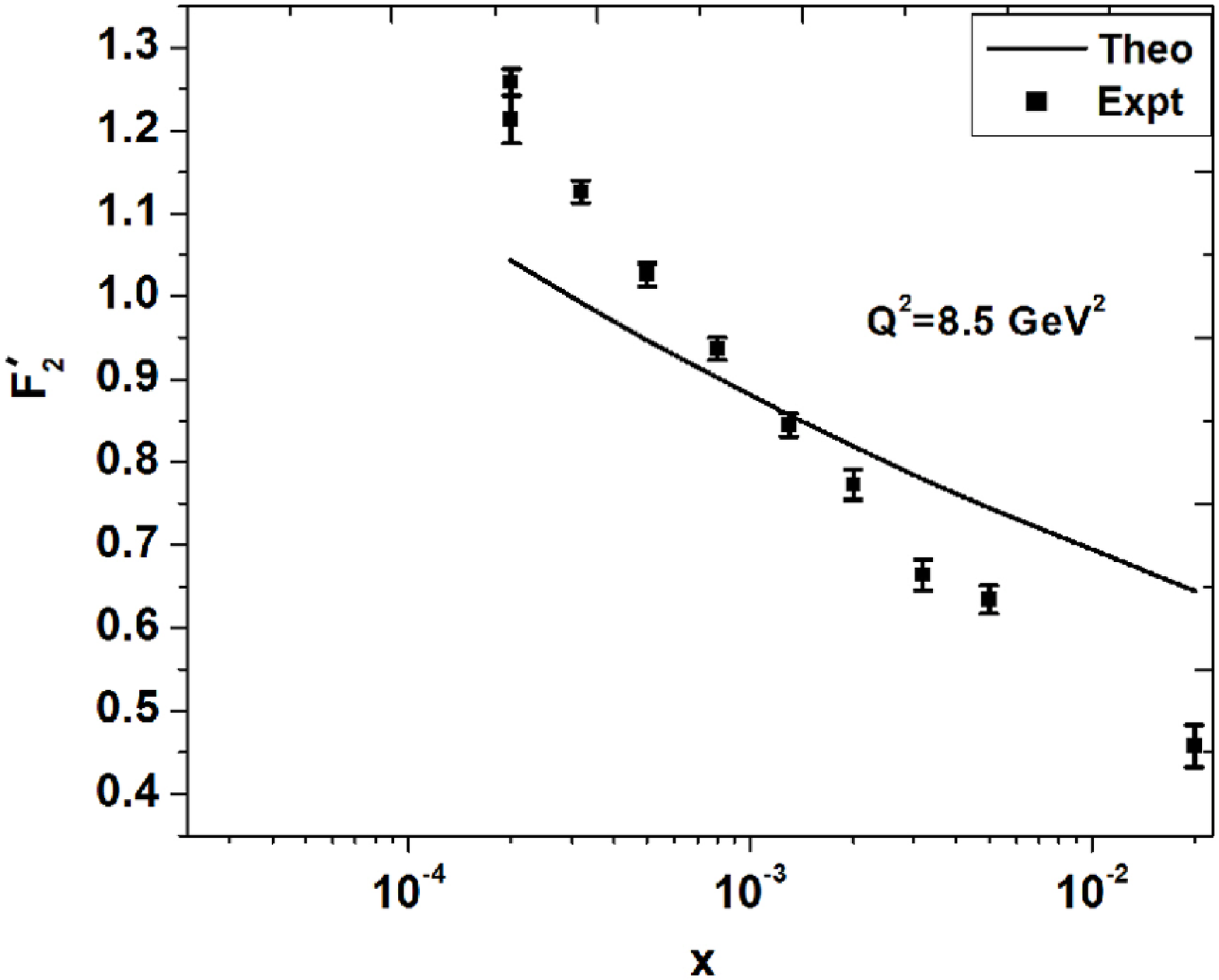}}\quad
 \subfloat[]{\includegraphics[width=.3\textwidth]{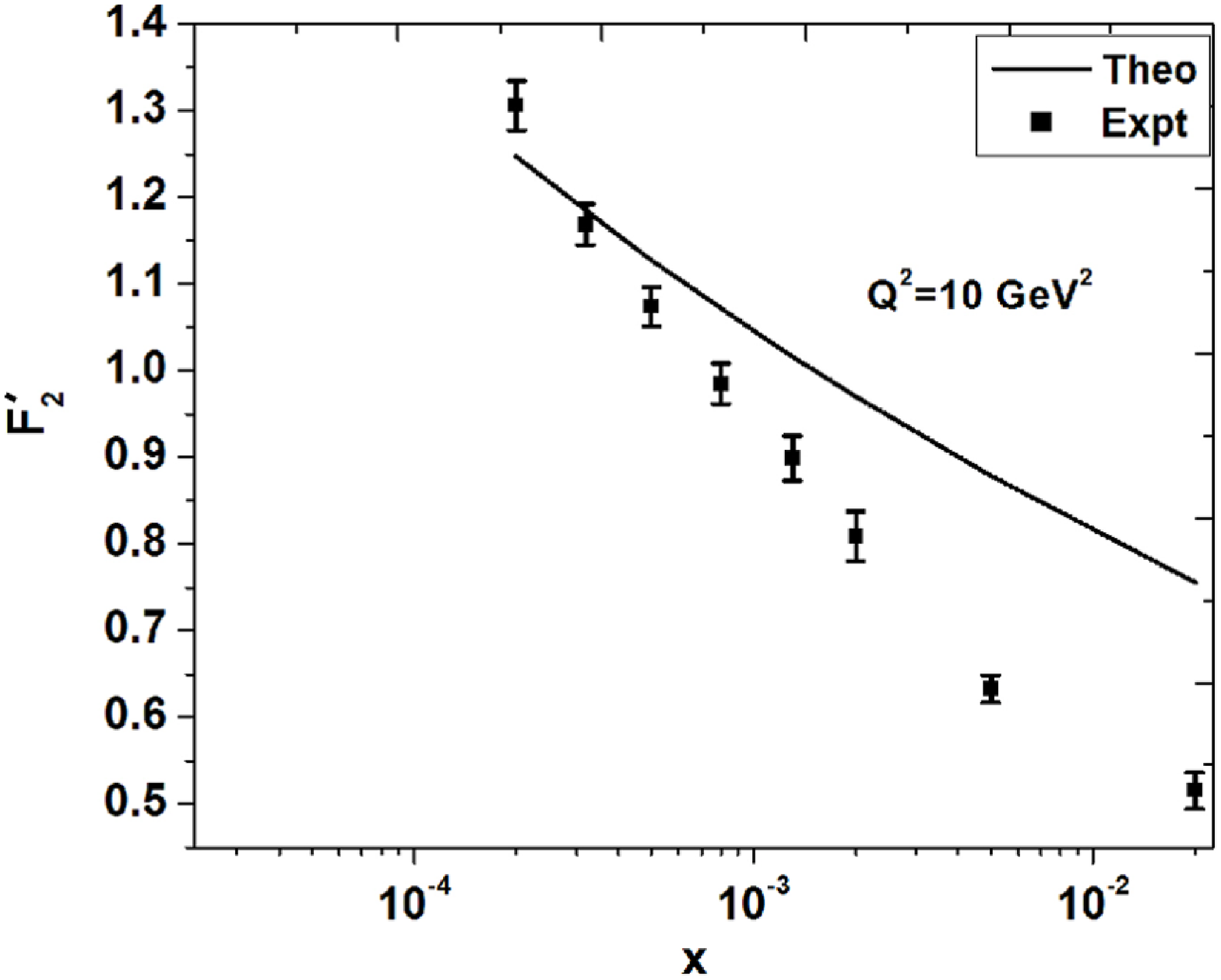}}\quad
\caption{{\footnotesize comparison of the structure function $F'_2$ of Model 2 as a function of $x$ in bins of $Q^2$ with measured data of $F_2$  from HERAPDF1.0\cite{HERA}}}
 \label{Fig1}
\end{figure*}

In Fig \ref{Fig1}, we plot $F'_2$ of Model 2 as a function of $x$ for six representative values of $Q^2$ ($Q^2$= 1.5, 2.7, 3.5, 6.5, 8.5, 10 GeV$^2$) in the phenomenologically allowed range 0.85 $\leq Q^2 \leq$ 10 GeV$^2$. We also show the corresponding available data from Ref\cite{HERA}.

It shows that the model parameters have more restrictive constraints. Due to positivity, the range of validity shrinks to $Q^2\leqslant 10$ GeV$^2$ from $Q^2\leqslant 120$ GeV$^2$. Thus our analysis indicates that the phenomenological range of validity of the present version (Model 2) is more restrictive than the previous version (Model 1).\\

\begin{table}[!bp]
\caption{\label{Table4}%
Results of the fit of Model 2}
\begin{ruledtabular}
\begin{tabular}{ccccccc}
\textrm{$D'_0$}&
\textrm{$D'_1$}&
\textrm{$D'_2$}&
\textrm{$D'_3$}&
\textrm{$Q_0'^2$(GeV$^2$)}&
\textrm{$\chi^2$}&
\textrm{$\chi^2$/ndf} \\
\colrule
-2.971\tiny${\pm 0.409}$ & 0.065\tiny${\pm 0.0003}$ & 1.021\tiny${\pm 0.004}$ & 0.0003\tiny${\pm 0.0001}$ & 0.20\tiny${\pm 0.0008}$ & 18.829 & 0.20\\
\end{tabular}
\end{ruledtabular}
\end{table}
We also observe the following features of the model compared to data: at $Q^2=1.5$GeV$^2$ data overshoots the theory. But as $Q^2$ increases, the theoretical curve comes closer to data. At $Q^2$=10 GeV$^2$, on the other hand, the theory exceeds data. Main reason of this feature is that the $x$-slope of the model is less than that of the data. Specifically, due to positive $D_3$, the growth of the structure function with $Q^2$ becomes faster  seen from Eqn(\ref{a2}) i.e.
\begin{equation}
\left( 1+\frac{Q^2}{Q_0'^2}\right) ^{(1+D'_3)} \approx \left( 1+\frac{Q^2}{Q_0'^2}\right) ^{1.0003}
\end{equation}
at higher values of $Q^2 >$ 1 GeV$^2$ to be compared with 
\begin{equation}
\left( 1+\frac{Q^2}{Q_0^2}\right) ^{(1+D_3)} \approx \left( 1+\frac{Q^2}{Q_0^2}\right) ^{-0.287}
\end{equation}
that of Ref\cite{Last}. This is the major limitation of the present singularity free version of the model: effort to make it singularity free reduces its phenomenological range of validity drastically.

\subsection{Graphical representation of TMD}
\label{3b}
It is interesting to predict the $k_t^2$-dependance of unintegrated parton distributions (TMD) from the $x$ and $Q^2$ dependence of the integrated parton distribution function (PDF). Clearly this can be done within a model framework, as has been noted in Ref\cite{ZA} as well as in Ref\cite{DK9}. Though it should be of interest to explore this approach to study $k_t^2$-dependence of $f_i(x,k_t^2)$, such a study makes sense only in the $x$-$Q^2$ range where the approach works and where the parameters had been fitted. Here we take the two models (Models 1 and 2) to analyze the TMDs graphically and compare their pattern with respect to \textit{x} and $k_t^2$.

\begin{figure}[!tbp]
\subfloat[]{%
  \includegraphics[width=0.49\linewidth]{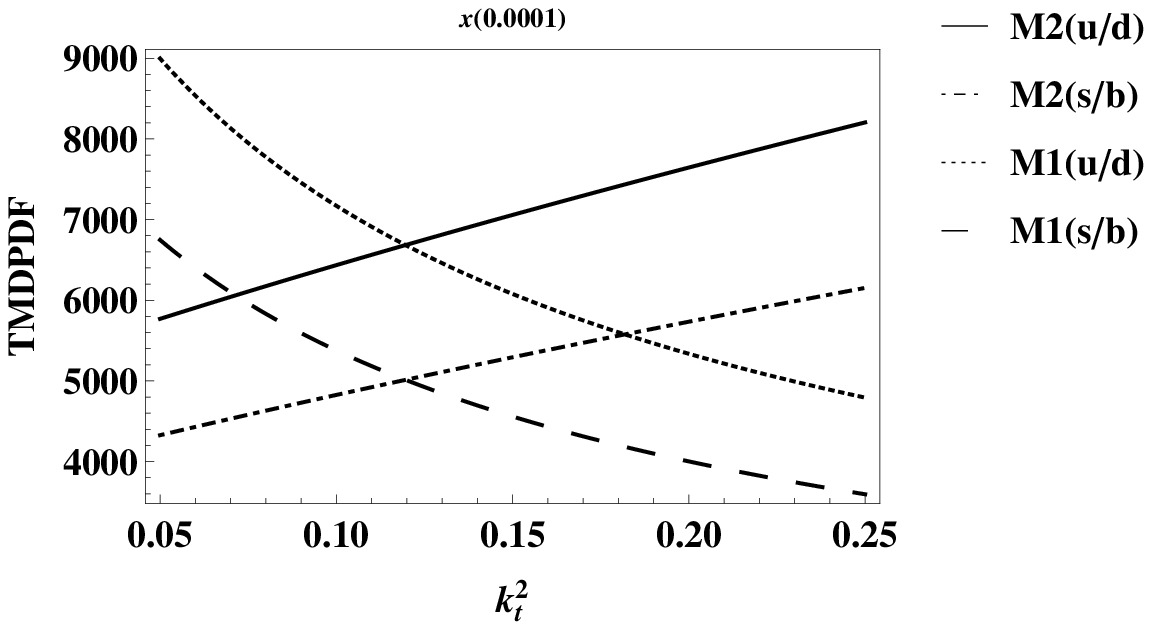}%
}\vspace{1ex}
\subfloat[]{%
  \includegraphics[width=0.47\linewidth]{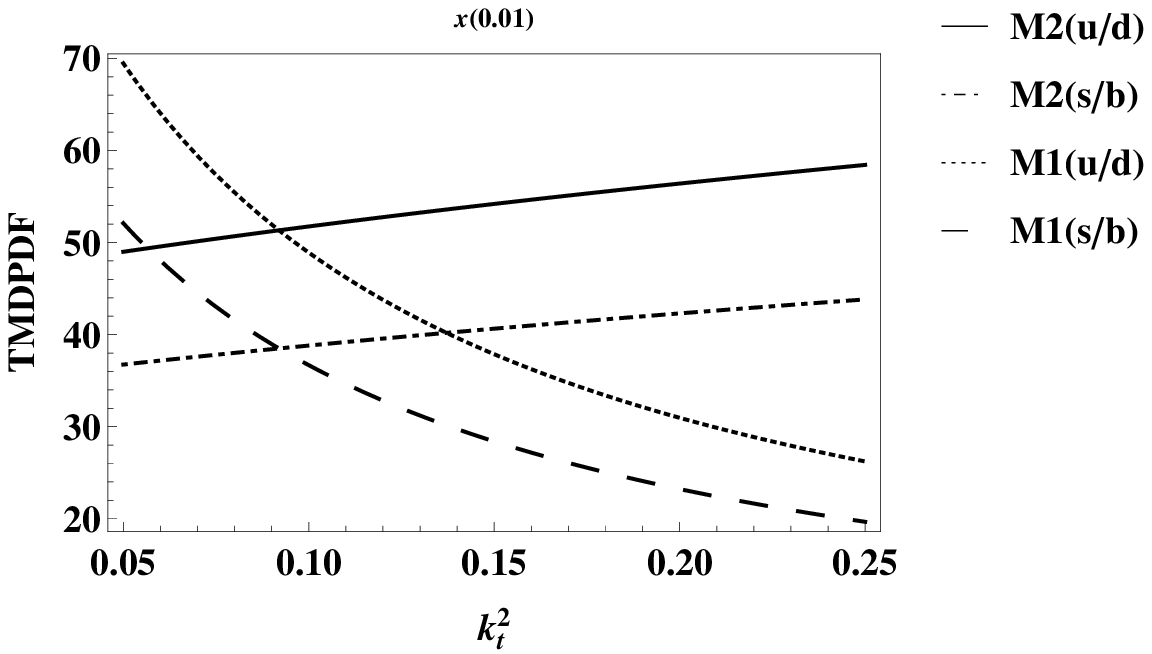}%
}\vspace{1ex}
\caption{TMDPDF vs $k_t^2$ for two representative values of (a) $x=10^{-4}$ and (b) $x=0.01$ for Models 1 and 2. Here, M1(u/d) (black dotted) and M2(u/d) (black line) represents the TMDPDF for u and d quarks for Models 1 and 2 respectively. Similarly, M1(s/b) (black dashed) and M2(s/b) (black dot-dashed) represents the TMDPDF for s and b quarks for Models 1 and 2 respectively.}
\label{F2}
\end{figure}

Using Eqn(\ref{E1}), TMDs for the two models can be written as:
\begin{eqnarray}
\label{E21}
{\text M\text o\text d\text e\text l \ \text 1:} \quad \quad \quad \quad & & f_i(x,k_t^2)=\frac{e^{D_0^i}}{M^2} \left(\frac{1}{x}\right)^{D_2+D_1 \log\left(1+\frac{k_t^2}{k_0^2}\right)} \left(1+\frac{k_t^2}{k_0^2}\right)^{D_3} \\
\label{E2a}
{\text M\text o\text d\text e\text l \ \text 2:} \quad \quad \quad \quad & & f'_i(x,k_t^2)=\frac{e^{D_0'^i}}{M^2} \left(\frac{1}{x}\right)^{D'_2+D'_1 \log\left(1+\frac{k_t^2}{k_0'^2}\right)} \left(1+\frac{k_t^2}{k_0'^2}\right)^{D'_3}
\end{eqnarray}
\\
To evaluate Eqn(\ref{E21}) and (\ref{E2a}), we take the mean value of the parameters from Eqn(\ref{E7}) as well as Table \ref{Table4} respectively. We use Eqn(8) with $n_f=4$ and assume u , d , s, and b are in the ratio 
\begin{eqnarray}
& e^{D_0^u} : e^{D_0^d} : e^{D_0^s} : e^{D_0^b} \nonumber \\
&  4 : 4 : 1 : 1
\end{eqnarray}
for definiteness. This gives $e^{{D_0}^u}=1.148=e^{{D_0}^d}$ and $e^{{D_0}^s}=0.287=e^{{D_0}^b}$. Similarly $e^{{D_0'}^u}=0.0416=e^{{D_0'}^d}$ and $e^{{D_0'}^s}=0.0104=e^{{D_0'}^b}$\\ In Fig \ref{F2}, TMDPDF vs $k_t^2$ is shown using Eqn(\ref{E21}) and (\ref{E2a}) for representative values of

\begin{figure}[!tbp]
\subfloat[]{%
  \includegraphics[width=0.48\linewidth]{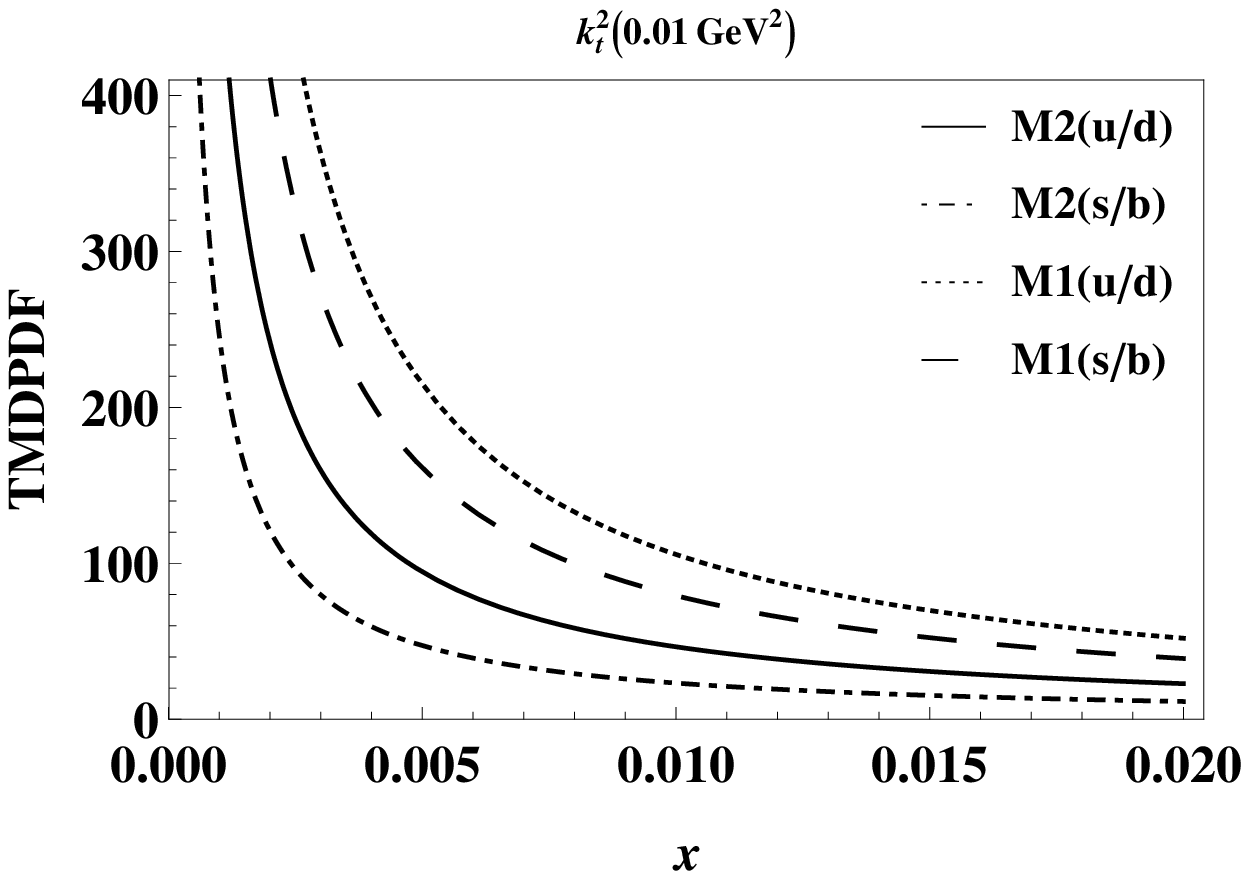}%
}\vspace{1ex}
\subfloat[]{%
  \includegraphics[width=0.48\linewidth]{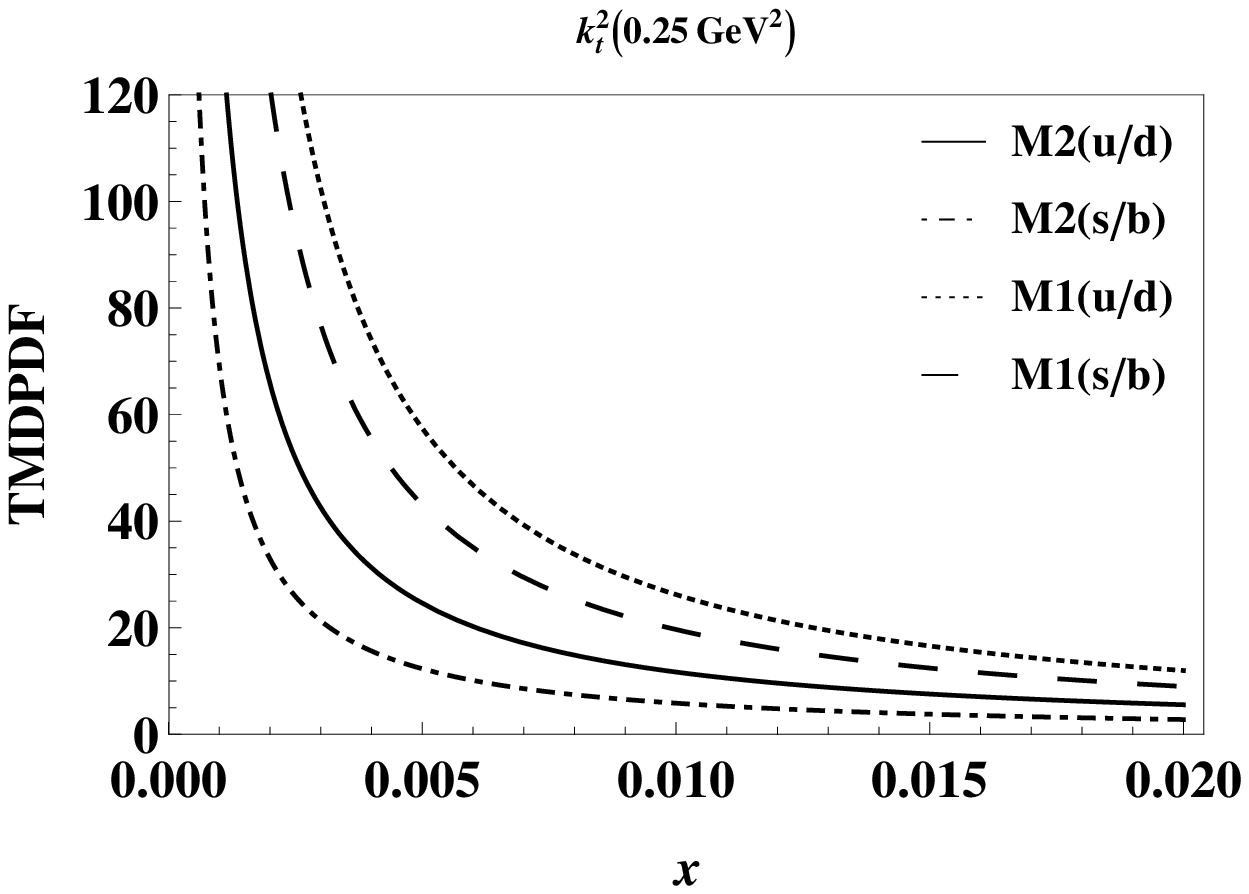}%
}\vspace{1ex}

\caption{TMDPDF vs \textit{x} for two representative values of (a) $k_t^2= 0.01$ GeV$^2$ and (b) $k_t^2= 0.25$ GeV$^2$ for Models 1 and 2. Here, M1(u/d) (black dotted) and M2(u/d) (black line) represents the TMDPDF for u and d quarks for Models 1 and 2 respectively. Similarly, M1(s/b) (black dashed) and M2(s/b) (black dot-dashed) represents the TMDPDF for s and b quarks for Models 1 and 2 respectively.}
\label{F3}
\end{figure}

(i) $x$ = 10$^{-4}$ and (ii) $x$=0.01 setting $M^2=1$ GeV$^2$.\\
The allowed limit of $k_t^2$ is considered to be less than the average value $k_t^2$=0.25 GeV$^2$ \cite{h2} as determined from data. Similarly TMDPDF vs $x$ is shown in Fig \ref{F3} for representative values of

(i) $k_t^2$ = 0.01 GeV$^2$ and (ii) $k_t^2$=0.25 GeV$^2$. 

The present graphical analysis of TMDs (Fig \ref{F2}-\ref{F3}) is a comparison of both the versions of self-similar models of proton structure function. From Fig \ref{F2}, it can be seen the singularity free version (Model 2) increases with increasing $k_t^2$ unlike the Model 1. This conflicting  behavior in TMDs is due to the parameter $D_3$ which is made positive to avoid the singularity present in the Model 1. It indicates that only if the structure function has a singularity in \textit{x} (which might be situated even outside the phenomenological range of validity) one can get the usual qualitative feature of TMD \cite{ZA} which decreases as $k_t^2$ increases. Thus to keep the TMD well behaved, negative $D_3$ appears to be a necessary condition which we will attempt to rectify later.\\
\\
Let us now compare the structure of the model TMDs (Eqn\ref{E21}-\ref{E2a}) with the suggested forms \cite{ZA,34} available in current literature.

The standard way to study TMDs is through the factorization approach \cite{h2,h,h1} where $x$ and $Q^2$-dependence are factorized into a PDF $q_i(x,Q^2)$ and a Gaussian transverse momentum dependent function $h(k_t^2)$ 
\begin{equation}
\label{E23}
f_i(x,k_t^2,Q^2)= q'_i(x,Q^2) h(k_t^2)
\end{equation}
where 
\begin{equation}
\label{E24}
h(k_t^2)= \frac{1}{\langle k_t^2\rangle} e^{-\frac{k_t^2}{\langle k_t^2\rangle}}
\end{equation}
with normalization condition 
\begin{equation}
\label{E25}
\int h(k_t^2)dk_t^2= 1
\end{equation}\\
Such factorization property of TMD is not present in the Models 1 and 2 (Eqn \ref{E21}-\ref{E2a}) nor the Gaussian form (Eqn \ref{E24}) \cite{ZA}. In this sense the present models are close to the corresponding non-factorisable models of Ref\cite{holo,33,34}. Only in the absence of correlation term $D_1$ (Eqn \ref{E1}) such factorization property emerges. In this limit the $k_t^2$ dependent functional form of TMD (regarding Models 1 and 2: Eqn \ref{E21}-\ref{E2a}) are given by 
\begin{equation}
\label{E26}
\tilde{h}(k_t^2)= \frac{1}{M^2}\left(1+\frac{k_t^2}{k_0^2}\right)^{D_3}
\end{equation}
and
\begin{equation}
\label{E28}
h'(k_t^2)= \frac{1}{M^2}\left(1+\frac{k_t^2}{k_0'^2}\right)^{D'_3}
\end{equation}
\\
respectively in contrast to a Gaussian function (Eqn \ref{E24}). Introducing a $k_t^2$ cut off \ $0<k_t^2<\langle k_t^2\rangle$ with $\langle k_t^2\rangle$= 0.25 GeV$^2$, Eqn (\ref{E26}) and (\ref{E28}) will satisfy the normalization condition (Eqn \ref{E25}) with a normalization constant 
\begin{equation}
\label{E27}
\tilde{N}= \frac{M^2 (D_3+1)}{k_0^2\left[\left(1+\frac{\langle k_t^2\rangle}{k_0^2}\right)^{D_3+1}-1\right]} 
\end{equation}
\begin{equation}
\label{E29}
N'= \frac{M^2 (D'_3+1)}{k_0'^2\left[\left(1+\frac{\langle k_t^2\rangle}{k_0'^2}\right)^{D'_3+1}-1\right]} 
\end{equation}
\begin{figure}[!tbp]
\centering
\includegraphics*[width=80mm]{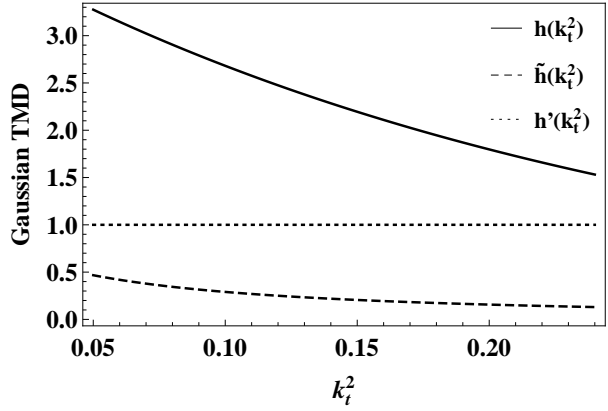}
\caption{{\small Gaussian TMD vs $k_t^2$} for $h(k_t^2)$ (black line) Eqn \ref{E24},\ $\tilde{h}(k_t^2)$ (black dashed) (Model 1),\ and $h'(k_t^2)$ (black dotted) (Model 2) respectively.}
\label{Fig5}
\end{figure}
respectively. In Fig \ref{Fig5}, we compare the Gaussian TMD Eqn(\ref{E24}) with the model TMDs (Eqn \ref{E26}-\ref{E28}) in the absence of the correlation term. We note that here the $k_t^2$-dependence is flavor independent. The qualitative feature of Fig \ref{Fig5} is identical to that of Fig \ref{F2}.

\section{Improved version of the self-similarity based models}
\label{D}
\subsection{A plausible way of removing the TMDPDF and PDF anomalies}
Let us discuss a possible way of removing the short coming of the models under discussion. As noted in Ref\cite{dkc} this approach has taken the notion of self-similarity to parametrize Parton Distribution Function (PDF) and eventually the structure function. However, the variables in which the supposed fractal scaling of the quark distributions and $F_2(x,Q^2)$ occur are not known from the underlying theory. In Ref\cite{Last}, the choice of $\frac{1}{x}$ is presumably because of the power law form of the quark distributions at small \textit{x} found in Gl{\"u}ck-Reya-Vogt (GRV)\cite{grv} distribution. However, this form is not derived theoretically but rather follows from the power law distributions in \textit{x} assumed for the input quark distributions used by the GRV distribution for the QCD evolution. The choice of $\frac{1}{x}$ as the proper scaling variable is therefore not established from the underlying theory. Instead, if $\log\frac{1}{x}$ is chosen as the scaling variable then asymptotic Froissart saturation like \cite{fe} behavior can be achieved in such self-similar model as in the QCD based model of Ref\cite{blo}. Same is true for the magnification factor $M_1=\left( 1+\frac{k_t^2}{k_0^2} \right)$ as occurred in defining TMD (Eqn \ref{E1}) which is the source of the anomalous relationship between the apparent necessity of a singularity of the structure function and the physically expected behavior of TMD to describe the finite intrinsic transverse momentum of partons in Proton.

The magnification factor $M_1$ can be considered as special case of a more general form :
\begin{equation}
\label{E30}
\hat{M}_1=\sum_{i=-n}^{n} \alpha_i M_1^i
\end{equation}
The qualitative feature of physically plausible TMDs can be achieved only if all the coefficients $\alpha _i (i=0, \ 1,\ 2, \ . \ . \ . \ , n)$ vanish. Only in a specific case, where $\alpha _1=1$ and all other coefficients cases vanish lead to the original $M_1$ as defined in Eqn(\ref{E1}). 

The defining TMD therefore can be generalized to
\begin{equation}
\label{E31}
\log[M^2.\hat{f}_i(x,k_t^2)]=\hat{D}_1\log\frac{1}{x}\log\hat{M}_1+\hat{D}_2\log\frac{1}{x}+\hat{D}_3\log\hat{M}_1+\hat{D}_0^i
\end{equation}
\\
instead of Eqn(\ref{E1}), such that the generalized TMD will take the form
\begin{equation}
\label{E32}
\hat{f}_i(x,k_t^2)=\frac{e^{\hat{D}_0^i}}{M^2}\left( \frac{1}{x}\right) ^{\hat{D}_2}\left( \hat{M}_1\right) ^{\hat{D}_3+\hat{D}_1\log\frac{1}{x}}
\end{equation}
\\
with the form of $\hat{M}_1$
\begin{equation}
\label{E33}
\hat{M}_1=\sum_{j=1}^{n} \frac{B_j}{\left( 1+\frac{k_t^2}{\hat{k}_0^2}\right) ^j}
\end{equation}
\\
where
\begin{equation}
B_j= \alpha_{-j}
\end{equation}
Taking only the two terms of Eqn(\ref{E33}), $\hat{M_1}$ can be written as
\begin{equation}
\label{E34}
\hat{M}_1=\frac{B_1}{\left( 1+\frac{k_t^2}{\hat{k}_0^2}\right)}+\frac{B_2}{\left( 1+\frac{k_t^2}{\hat{k}_0^2}\right) ^2}
\end{equation}
\\
and the corresponding TMD (Eqn \ref{E32}) becomes
\begin{equation}
\label{E35}
\hat{f}_i(x,k_t^2)=\frac{e^{\hat{D}_0^i}}{M^2}\left( \frac{1}{x} \right)^{\hat{D}_2}\left( \frac{B_1}{\left( 1+\frac{k_t^2}{\hat{k}_0^2} \right)} \right)^{\hat{D}_3+\hat{D}_1\log\frac{1}{x}} \left( 1+ \frac{B_2}{B_1} \frac{1}{\left( 1+ \frac{k_t^2}{\hat{k}_0^2} \right) }  \right)^{\hat{D}_3+\hat{D}_1\log\frac{1}{x}}
\end{equation}
\\
Assuming the convergence of the polynomials as occurred in Eqn(\ref{E35}) we obtain : \\ \textbf{(Model 3)}
\begin{equation}
\label{E36}
\hat{f}_i(x,k_t^2)=\frac{e^{\hat{D}_0^i}}{M^2}\left( \frac{1}{x} \right)^{\hat{D}_2}\left( \frac{B_1}{\left( 1+\frac{k_t^2}{\hat{k}_0^2} \right)} \right)^{\hat{D}_3+\hat{D}_1\log\frac{1}{x}} \left( 1+ \frac{B_2}{B_1} \frac{\left( \hat{D}_3+\hat{D}_1\log\frac{1}{x} \right) }{\left( 1+ \frac{k_t^2}{\hat{k}_0^2} \right) }  \right)
\end{equation}
\\
After integration over $k_t^2$, it yields the desired PDF
\begin{multline}
\label{E37}
\hat{q}_i(x,Q^2)=\frac{e^{\hat{D}_0^i} \hat{Q}_0^2}{M^2}\left( \frac{1}{x} \right)^{\hat{D}_2} \left( B_1 \right)^{\left( \hat{D}_3+\hat{D}_1\log\frac{1}{x} \right)}
\\
\left[  \frac{\left( \left(1+\frac{Q^2}{\hat{Q}_0^2} \right)^{\left( 1-\hat{D}_3-\hat{D}_1\log\frac{1}{x}\right) } -1 \right) }{\left( 1-\hat{D}_3-\hat{D}_1\log\frac{1}{x}\right) } -\frac{B_2}{B_1}\left(  \left(1+\frac{Q^2}{\hat{Q}_0^2} \right)^{\left( -\hat{D}_3-\hat{D}_1\log\frac{1}{x}\right) } -1 \right) \right] 
\end{multline}
\\
Using Eqn(\ref{E37}) in Eqn(\ref{E5}), the usual definition of structure function, it gives 
\begin{multline}
\label{E38}
\hat{F}_2(x,Q^2)=\frac{e^{\hat{D}_0} \hat{Q}_0^2}{M^2}\left( \frac{1}{x} \right)^{\hat{D}_2-1} \left( B_1 \right)^{\left( \hat{D}_3+\hat{D}_1\log\frac{1}{x} \right)}
\\
\left[  \frac{\left( \left(1+\frac{Q^2}{\hat{Q}_0^2} \right)^{\left( 1-\hat{D}_3-\hat{D}_1\log\frac{1}{x}\right) } -1 \right) }{\left( 1-\hat{D}_3-\hat{D}_1\log\frac{1}{x}\right) } -\frac{B_2}{B_1}\left( \left(1+\frac{Q^2}{\hat{Q}_0^2} \right)^{\left( -\hat{D}_3-\hat{D}_1\log\frac{1}{x}\right) } -1 \right) \right] 
\end{multline}
\\
with the condition that 
\begin{equation}
\hat{D}_3+\hat{D}_1\log\frac{1}{x}\neq1
\end{equation}
as the equality will yield a undesired singularity. \\
The above model of structure function (Model 3) has new 7 independent parameters $B_1,\ B_2,\ \hat{D}_0,\ \hat{D}_1, \ \hat{D}_2,\ \hat{D}_3,\ \hat{Q}_0^2$ to be fitted from data and compared with the previous models (Models 1 and 2). If the model parameters $\hat{D}_1$ and $\hat{D}_3$ satisfy the additional condition
\begin{equation}
\label{1}
\hat{D}_3+\hat{D}_1\log\frac{1}{x}=1
\end{equation}
then the resultant TMD becomes : \textbf{(Model 4)}
\begin{equation}
\label{E39}
\tilde{f}_i(x,k_t^2)=\frac{e^{\tilde{D}_0^i}}{M^2}\left( \frac{1}{x} \right)^{\tilde{D}_2}\left( \frac{\tilde{B}_1}{\left( 1+\frac{k_t^2}{\tilde{k}_0^2} \right)} \right) \left( 1+ \frac{\tilde{B}_2}{\tilde{B}_1} \frac{1}{\left( 1+ \frac{k_t^2}{\tilde{k}_0^2} \right) }  \right)
\end{equation}
\\
while the integration over $k_t^2$ leads to the PDF
\begin{equation}
\label{E40}
\tilde{q}_i(x,Q^2)=\frac{e^{\tilde{D}_0^i} \tilde{Q}_0^2}{M^2}\left( \frac{1}{x} \right)^{\tilde{D}_2} \tilde{B}_1 \left[ \log \left( 1+\frac{Q^2}{\tilde{Q}_0^2} \right) -\frac{\tilde{B}_2}{\tilde{B}_1}\left(\frac{1}{\left( 1+\frac{Q^2}{\tilde{Q}_0^2}\right) }-1 \right)  \right] 
\end{equation}
\\
And the corresponding structure function is
\begin{equation}
\label{E41}
\tilde{F}_2(x,Q^2)=\frac{e^{\tilde{D}_0} \tilde{Q}_0^2}{M^2}\left( \frac{1}{x} \right)^{\tilde{D}_2-1} \tilde{B}_1 \left[ \log \left( 1+\frac{Q^2}{\tilde{Q}_0^2} \right) -\frac{\tilde{B}_2}{\tilde{B}_1}\left(\frac{1}{\left( 1+\frac{Q^2}{\tilde{Q}_0^2}\right) }-1 \right)  \right] 
\end{equation}
\\
which is completely free from singularity except for $\tilde{D}_2\geq1$ . Such singularity is, however, consistent with the usual Regge expectation \cite{reg,m,boo,coo,yu}. The model has now got 4 parameters: $\tilde{B}_1, \ \tilde{D}_2, \ \tilde{Q}_0^2, \ \tilde{D}_0^i$.

\subsection{Extrapolation of the self-similarity based TMDPDF from small \textit{x} to large \textit{x}}
The models of TMDPDFs or PDFs discussed above were basically constructed to test it in the small \textit{x} range. It did not take into account the large \textit{x} behavior \cite{bro,m,boo,coo,yu} of the PDF or structure function
\begin{equation}
\label{E43}
\lim_{x\to1} F_2(x,Q^2)=0
\end{equation}
which is not unexpected. The important observation which motivated and justified the use of self-similarity concept was that for $x < 0.01$; the logarithm of the derivative of the unintegrated parton distributions $\log\left( \frac{\partial f_i( x,Q^2 )}{\partial Q^2}\right)$ is a linear function of $\log x$ (Fig 2.8.a of Ref\cite{Last}). The idea of self-similarity is based on the fact that at small \textit{x}, the behavior of quark density is driven by gluon emissions and splittings such that the parton distribution function at small \textit{x} and those at still smaller \textit{x} look similar (upto some magnification factor). In the opposite limit, at large \textit{x}, there is no physical reason for self-similarity and no phenomenological justification till date. In other words, extending the approach of large $x$ means applying the self-similarity concept where it is not expected to work. On the other hand, it is not unreasonable to assume that the self-similarity does not terminate abruptly at $x \approx 0.01$, but smoothly vanishes at $x = 1$, the valence quark limit of proton with no trace of self-similarity at all.

As in Ref\cite{DK6}, we take this alternative point of view in structure function. We suggest a simple interpolating models TMDPDF/PDF which approaches the self-similar one at $x \rightarrow 0$ (Eqn \ref{E1}), and still satisfy Eqn(\ref{E43}) at large \textit{x}, $x \rightarrow 1$. A plausible way of achieving it in a parameter-free way is to make a formal replacement of $\frac{1}{x}$ factor to $\left(\frac{1}{x}-1\right)$ in Eqn(\ref{E1}). The former one is identified as one of the magnification factors in the self-similar model, while the later can be so interpreted only for $\frac{1}{x}\gg 1$. In such case, Eqn(\ref{E1}) will be modified to $\bar{f}_i(x, k_t^2 )$ defined as
\begin{equation}
\label{E44}
\log[M^2.\bar{f}_i(x,k_t^2)]= \bar{D}_1.\log\left( \frac{1}{x}-1\right).\log\left(1+\frac{k_t^2}{\bar{k}_0^2}\right)+\bar{D}_2.\log\left( \frac{1}{x}-1\right) +\bar{D}_3.\log\left(1+\frac{k_t^2}{\bar{k}_0^2}\right)+\bar{D}_0^i
\end{equation}
which leads to 
\begin{equation}
\bar{f}_i(x,k_t^2)= \frac{e^{\bar{D}_0^i}}{M^2} \left( \frac{1}{x}-1\right)^{D_2}\left(1+\frac{k_t^2}{\bar{k}_0^2}\right)^{D_3+D_1\log\left( \frac{1}{x}-1\right)}
\end{equation}
Generalizing the magnification factor $\hat{M_1}$ as in Eqn(\ref{E34}) and taking only the two terms and assuming the convergence of the polynomials occurring in the expression as in Eqn(\ref{E35}) we obtain the generalized TMD as : \textbf{(Model 5)}
\begin{multline}
\label{E45}
\bar{f}_i(x,k_t^2)=\frac{e^{\bar{D}_0^i}}{M^2}\left( \frac{1}{x} \right)^{\bar{D}_2}\left(1-x \right)^{\bar{D}_2}\left( \frac{\bar{B}_1}{\left( 1+\frac{k_t^2}{k_0^2} \right)} \right)^{\bar{D}_3+\bar{D}_1\log\frac{1}{x}+\bar{D}_1\log(1-x)} 
\\
\left( 1+ \frac{\bar{B}_2}{\bar{B}_1} \frac{\left( \bar{D}_3+\bar{D}_1\log\frac{1}{x}+\bar{D}_1\log(1-x) \right) }{\left( 1+ \frac{k_t^2}{\bar{k}_0^2} \right) }  \right)
\end{multline}
\\
And hence corresponding PDF$\left(\bar{q}_i\right)$  and structure function$\left( \bar{F}_2\right) $ will be
\begin{multline}
\label{E44a}
\bar{q}_i(x,Q^2)=\frac{e^{\bar{D}_0^i} \bar{Q}_0^2}{M^2}\left( \frac{1}{x} \right)^{\bar{D}_2}\left(1-x \right)^{\bar{D}_2}  \left( \bar{B}_1 \right)^{\left( \bar{D}_3+\bar{D}_1\log\frac{1}{x}+\bar{D}_1\log(1-x)\right)}
\\
\left[  \frac{\left( \left(1+\frac{Q^2}{\bar{Q}_0^2} \right)^{\left( 1-\bar{D}_3-\bar{D}_1\log\frac{1}{x}-\bar{D}_1\log(1-x)\right) } -1 \right) }{\left( 1-\bar{D}_3-\bar{D}_1\log\frac{1}{x}-\bar{D}_1\log(1-x)\right) } -\frac{\bar{B}_2}{\bar{B}_1}\left( \left(1+\frac{Q^2}{\bar{Q}_0^2} \right)^{\left( -\bar{D}_3-\bar{D}_1\log\frac{1}{x}-\bar{D}_1\log(1-x)\right) } -1 \right) \right] 
\end{multline}
and 
\begin{multline}
\label{E45a}
\bar{F}_2(x,Q^2)=\frac{e^{\bar{D}_0} \bar{Q}_0^2}{M^2}\left( \frac{1}{x} \right)^{\bar{D}_2-1} \left(1-x \right)^{\bar{D}_2-1}  \left( \bar{B}_1 \right)^{\left( \bar{D}_3+\bar{D}_1\log\frac{1}{x}+\bar{D}_1\log(1-x)\right)}
\\
\left[  \frac{\left( \left(1+\frac{Q^2}{\bar{Q}_0^2} \right)^{\left( 1-\bar{D}_3-\bar{D}_1\log\frac{1}{x}-\bar{D}_1\log(1-x)\right) } -1 \right) }{\left( 1-\bar{D}_3-\bar{D}_1\log\frac{1}{x}-\bar{D}_1\log(1-x)\right) } -\frac{\bar{B}_2}{\bar{B}_1}\left( \left(1+\frac{Q^2}{\bar{Q}_0^2} \right)^{\left( -\bar{D}_3-\bar{D}_1\log\frac{1}{x}-\bar{D}_1\log(1-x)\right) } -1 \right) \right] 
\end{multline}
Imposing the condition 
\begin{equation}
\label{2}
\bar{D}_3+\bar{D}_1\log\frac{1}{x}+\bar{D}_1\log(1-x)=1
\end{equation}
will lead to corresponding TMD, PDF and structure function as : \textbf{(Model 6)}
\begin{equation}
\label{E46}
\bar{f}_i'(x,k_t^2)=\frac{e^{\bar{D}_0'^i}}{M^2}\left( \frac{1}{x} \right)^{\bar{D}'_2}\left(1-x \right)^{\bar{D}'_2}\left( \frac{\bar{B}'_1}{\left( 1+\frac{k_t^2}{k_0'^2} \right)} \right) \left( 1+ \frac{\bar{B}'_2}{\bar{B}'_1} \frac{1}{\left( 1+ \frac{k_t^2}{\bar{k}_0'^2} \right) }  \right)
\end{equation}
Corresponding PDF
\begin{equation}
\label{E47}
\bar{q}'_i(x,Q^2)=\frac{e^{\bar{D}_0'^i} \bar{Q}_0'^2}{M^2}\left( \frac{1}{x} \right)^{\bar{D}'_2} \left(1-x \right)^{\bar{D}'_2} \bar{B}'_1 \left[ \log \left( 1+\frac{Q^2}{\bar{Q}_0'^2} \right) -\frac{\bar{B}'_2}{\bar{B}'_1}\left(\frac{1}{\left( 1+\frac{Q^2}{\bar{Q}_0'^2}\right) }-1 \right)  \right] 
\end{equation}
and corresponding structure function
\begin{equation}
\label{E48}
\bar{F}'_2(x,Q^2)=\frac{e^{\bar{D}'_0} \bar{Q}_0'^2}{M^2}\left( \frac{1}{x} \right)^{\bar{D}'_2-1}\left(1-x \right)^{\bar{D}'_2} \bar{B}'_1 \left[ \log \left( 1+\frac{Q^2}{\bar{Q}_0'^2} \right) -\frac{\bar{B}'_2}{\bar{B}'_1}\left(\frac{1}{\left( 1+\frac{Q^2}{\bar{Q}_0'^2}\right) }-1 \right)  \right] 
\end{equation}

\subsection{Comparison of self-similarity PDF with standard PDF}
As in Ref\cite{DK6} we will now compare the parametrization of self-similarity PDF Eqn(\ref{E3}), (\ref{a1}), (\ref{E37}), (\ref{E40}), (\ref{E44a}), and (\ref{E47}) with the common behavior of quark and gluon distributions obtained in the standard parametrization like CTEQ \cite{ct}. Setting $Q^2=Q_0^2$ we have from Eqn(\ref{E3}), (\ref{a1}), (\ref{E37}), (\ref{E40}), (\ref{E44a}), and (\ref{E47}) \\ \\
\textbf{(Model 1)}
\begin{equation}
\label{E49}
q_i(x,Q_0^2)= A_1^i \left( \frac{1}{x}\right)^{D_2} \left( \left( \frac{1}{x}\right) ^{D_1\log2} \ 2^{D_3+1} -1 \right) 
\end{equation}
where 
\begin{equation}
A_1^i=\frac{e^{D_0^i} \ Q_0^2}{M^2 \ l_1(x)}
\end{equation}
Here
\begin{equation}
\label{l1}
l_1(x)= 1+D_3+D_1\log\frac{1}{x}
\end{equation}
\\
\textbf{(Model 2)}
\begin{equation}
\label{E49a}
q'_i(x,Q_0^2)=A_2^i \left( \frac{1}{x}\right)^{D'_2} \left( \left( \frac{1}{x}\right) ^{D'_1\log2} \ 2^{D'_3+1} -1 \right) 
\end{equation}
where 
\begin{equation}
A_2^i= \frac{e^{D_0'^i}\ Q_0'^2}{M^2 \ l_2(x)} 
\end{equation}
Here
\begin{equation}
\label{l2}
l_2(x)= 1+D'_3+D'_1\log\frac{1}{x}
\end{equation}
\\
\textbf{(Model 3)}
\begin{multline}
\label{E50}
\hat{q}_i(x,Q_0^2)= A_3^i  \left( \frac{1}{x}\right)^{\hat{D}_2} 
\\
\left( W_1 \left( \frac{1}{x}\right)^{\hat{D}_1\log\frac{B_1}{2}} + W_2 \left( \frac{1}{x}\right)^{\hat{D}_1\log B_1} + W_3 \log\frac{1}{x} \left( \frac{1}{x}\right)^{\hat{D}_1\log\frac{B_1}{2}}+ W_4 \log\frac{1}{x} \left( \frac{1}{x}\right)^{\hat{D}_1\log B_1} \right)  
\end{multline}
where 
\begin{equation}
A_3^i= \frac{e^{\hat{D}_0^i} \ \hat{Q}_0^2}{M^2 \ l_3(x)}
\end{equation}
Here
\begin{equation}
\label{l3}
l_3(x)= 1-\hat{D}_3-\hat{D}_1\log\frac{1}{x}
\end{equation}
And
\begin{eqnarray}
W_1 &=& B_1^{\hat{D}_3} 2^{1-\hat{D}_3}- 2^{-\hat{D}_3} B_1^{\hat{D}_3-1}\left(B_2-B_2 \hat{D}_3 \right)   \\
W_2 &=& (B_2- B_2 \hat{D}_3) B_1^{\hat{D}_3-1}- B_1^{\hat{D}_3} \\
W_3 &=& B_2 \hat{D}_1 2^{-\hat{D}_3} B_1^{\hat{D}_3-1} \\
W_4 &=& -B_2 \hat{D}_1 B_1^{\hat{D}_3-1}
\end{eqnarray}
\\
\textbf{(Model 4)}
\begin{equation}
\label{E40a}
\tilde{q}_i(x,Q_0^2)=A_4^i \left( \frac{1}{x}\right)^{\tilde{D}_2} \left(\tilde{B}_1 \log2 +\frac{1}{2} \tilde{B}_2 \right) 
\end{equation}
where
\begin{equation}
A_4^i= \frac{e^{\tilde{D}_0^i} \tilde{Q}_0^2}{M^2}
\end{equation}
\\
\textbf{(Model 5)}
\begin{footnotesize}
\begin{multline}
\label{E50a}
\bar{q}_i(x,Q_0^2)= A_5^i \left(\frac{1}{x}\right)^{\bar{D}_2}\ \left( 1-x \right)^{\bar{D}_2} 
\\
\left( Z_1 \left( \frac{1}{x}-1\right)^{\bar{D}_1\log\frac{\bar{B}_1}{2}} + Z_2 \left( \frac{1}{x}-1\right)^{\bar{D}_1\log\bar{B}_1} + Z_3 \log\left( \frac{1}{x} -1\right) \left( \frac{1}{x}-1\right)^{\bar{D}_1\log\frac{\bar{B}_1}{2}}+ Z_4 \log\left( \frac{1}{x}-1\right)  \left( \frac{1}{x}-1\right)^{\bar{D}_1\log\bar{B}_1} \right)
\end{multline}
\end{footnotesize} 
where 
\begin{equation}
A_5^i= \frac{e^{\bar{D}_0^i} \ \bar{Q}_0^2}{M^2\ l_4(x)}
\end{equation}
Here
\begin{equation}
\label{l4}
l_4(x)= 1-\bar{D}_3-\bar{D}_1\log\frac{1}{x}-\bar{D}_1\log(1-x)
\end{equation}
And
\begin{eqnarray}
Z_1 &=& \bar{B}_1^{\bar{D}_3} 2^{1-\bar{D}_3}- 2^{-\bar{D}_3} \bar{B}_1^{\bar{D}_3-1}\left(\bar{B}_2-\bar{B}_2 \bar{D}_3 \right) \\
Z_2 &=& (\bar{B}_2- \bar{B}_2 \bar{D}_3) \bar{B}_1^{\bar{D}_3-1}- \bar{B}_1^{\bar{D}_3} \\
Z_3 &=& \bar{B}_2 \bar{D}_1 2^{-\bar{D}_3} \bar{B}_1^{\bar{D}_3-1} \\
Z_4 &=& -\bar{B}_2 \bar{D}_1 \bar{B}_1^{\bar{D}_3-1}
\end{eqnarray}
\\
\textbf{(Model 6)}
\begin{equation}
\label{E51}
\bar{q}_i'(x,Q_0^2)=A_6^i \left( \frac{1}{x}\right)^{\bar{D}'_2} \left( 1-x \right)^{\bar{D}'_2} \left(  \bar{B}'_1 \log2 + \frac{1}{2} \bar{B}'_2 \right) 
\end{equation}
Where
\begin{equation}
A_6^i=\frac{e^{\bar{D}_0'^i} \ \bar{Q}_0'^2}{M^2} 
\end{equation}
\\
The \textit{x}-dependence of $l_1(x)$ and $l_2(x)$ defined above are due to the correlation between two magnification factors $M_1= \left( 1+\frac{k_t^2}{k_0^2}\right)$ and $M_2= \frac{1}{x}$ (Eqn \ref{E1}). Similarly \textit{x}-dependence of $l_3(x)$ and $l_4(x)$ are due to the correlation between $M_2$ and $\hat{M}_1$ (Eqn \ref{E31}). In Eqn(\ref{E40a}) and (\ref{E51}), the extra \textit{x}-dependence do not occur due to the initial conditions of logarithmic rise (Eqn \ref{1} and \ref{2}). If the terms occurring in $D_1$s are assumed to be negligible, then Eqn(\ref{E51}) has a form similar to the canonical parametrization \cite{boo,yu}
\begin{equation}
\label{E52}
q_i(x,Q_0^2) \approx A_0^i \ x^{A_1^i} \ (1-x)^{A_2^i}
\end{equation}
where the superscript \textit{i} indicates flavor dependence. At small \textit{x} it reduces to Eqn(\ref{E40a}). 

Let us construct the number of parameters as occurred in standard canonical parametrization and self-similarity parametrization Eqn(\ref{E3}), (\ref{a1}), (\ref{E37}), (\ref{E40}), (\ref{E44a}), and (\ref{E47}) . If $n_f$ is the number of flavors for both quarks and anti quarks then the number of parameters in Eqn(\ref{E52}) will be $6n_f+3$. The first factor is due to the quark and anti quark flavors and additional number 3 corresponding to the 3 parameters $A_0^q$ , $A_1^q$, and $A_2^q$ for gluon distributions. In a self-similar parametrization like Eqn(\ref{E49}-\ref{E51}), the exponents of \textit{x} and $B_1$s and  $B_2$s all are flavor independent. It implies, each flavor does not distinguish quark and anti quark. Thus the number of parameters in self-similar PDFs for the above models (1-6) are given in Table \ref{T2}. The first brackets in Column 2 of Table \ref{T2} correspond to number of parameters for quarks, while the second one, number of parameters for gluon.
\begin{table}[!tbp]
\caption{\label{T2}} {Number of parameters in self-similar pdf}
\begin{center}
\begin{tabular}{| c | c |}\hline
\ \ Models \ \ & \ \ Parameters \  \ \\ \hline
1 & $(n_f+4)+(4+1)$ \\ 
2 & $(n_f+4)+(4+1)$ \\
3 & $(n_f+6)+(6+1)$ \\
4 & $(n_f+4)+(4+1)$ \\
5 & $(n_f+6)+(6+1)$ \\
6 &  $(n_f+4)+(4+1)$ \\ \hline
\end{tabular}
\end{center}
\end{table}

The CTEQ \cite{ct}, more recent HERAPDF1.0 \cite{HERA}, HERAPDF2.0 \cite{pdf2}, and Ref\cite{m} parametrization have the corresponding forms
\begin{eqnarray}
\label{E53}
q_i^1(x,Q_0^2) &=& A_0^i x^{A_1^i}\left( 1-x \right)^{A_2^i} L_1(x) \\
\label{E54}
q_i^2(x,Q_0^2) &=& A^i x^{B^i} (1-x)^{C^i} L_2(x)  \\
\label{E55}
q_i^3(x,Q_0^2) &=& A^i x^{B^i} (1-x)^{C^i} L_3(x) \\
\label{E555}
q_i^4(x,Q_0^2) &=& A_{f_i} x^{a_{f_i}} (1-x)^{b_{f_i}} L_4(x)  
\end{eqnarray}
respectively, where
\begin{eqnarray}
L_1(x) &=& e^{A_3^i x} \left( 1 + e^{A_4^i}x \right)^{A_5^i} \\
L_2(x) &=& \left(1+ \epsilon \surd x + Dx + Ex^2 \right) \\
L_3(x) &=& \left(1 + Dx + Ex^2 \right) \\
L_4(x) &=& \mathscr{F} \left(x,\left\lbrace c_{f_i}\right\rbrace  \right) 
\end{eqnarray}
Here $L_4$ basically represents a smooth function which remains finite both $x\rightarrow 0$ and $x\rightarrow 1$. In the limit $D_1$s=0, the terms occurring in $D_1$s $(D_1, D'_1, \hat{D}_1, \bar{D}_1)$ becomes zero and the Eqn (\ref{E49}, \ref{E49a}, \ref{E50}, \ref{E50a}) have the similar form to the corresponding standard parametrization (Eqn \ref{E53}-\ref{E555}).
 
\subsection{Comparison of the structure function of Models 4 and 6 with data and determination of the corresponding TMDPDFs and PDFs:}
\subsubsection{\textbf{Comparison of data of Models 4 and 6 and determination of the model \\ parameters:}}
In this section, we make a comparison of TMDPDF/PDF and structure function of Models 4 and 6 since only these two have logarithmic $Q^2$ rise in PDF and structure function. Model 6 is the large \textit{x} extrapolation of Model 4.

To determine the parameters of Model 4 and Model 6, we have used the compiled HERA data \cite{HERA} as used in earlier work (Model 2). We make $\chi^2$-analysis of the data and obtained the phenomenological range of validity of $Q^2$ and \textit{x}.

For Model 4 the fitted parameters are given in Table \ref{T3} . The range of validity is found within $1.2\leq Q^2 \leq$ 800 GeV$^2$ and $2\times10^{-5}\leq x\leq 0.4$. The number of data points of $\tilde{F}_2$ is 284. Similarly for Model 6 the range of validity is $1.2\leq Q^2 \leq$ 1200 GeV$^2$ and $2\times10^{-5}\leq x\leq 0.4$ which is quite large in comparative to earlier works (Models 1 and 2). The fitted parameters for Model 6 are given in Table \ref{T4}. The number of data points of $\bar{F}'_2$ is 302.

\begin{figure*}[!tbp]
 \captionsetup[subfigure]{labelformat=empty}
\centering
  \subfloat[]{\includegraphics[width=.28\textwidth]{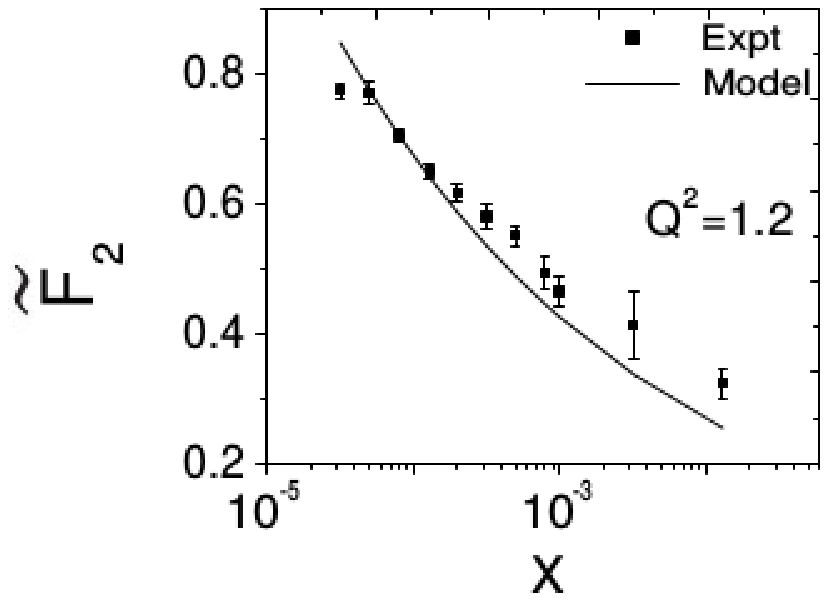}}\quad
 \subfloat[]{\includegraphics[width=.3\textwidth]{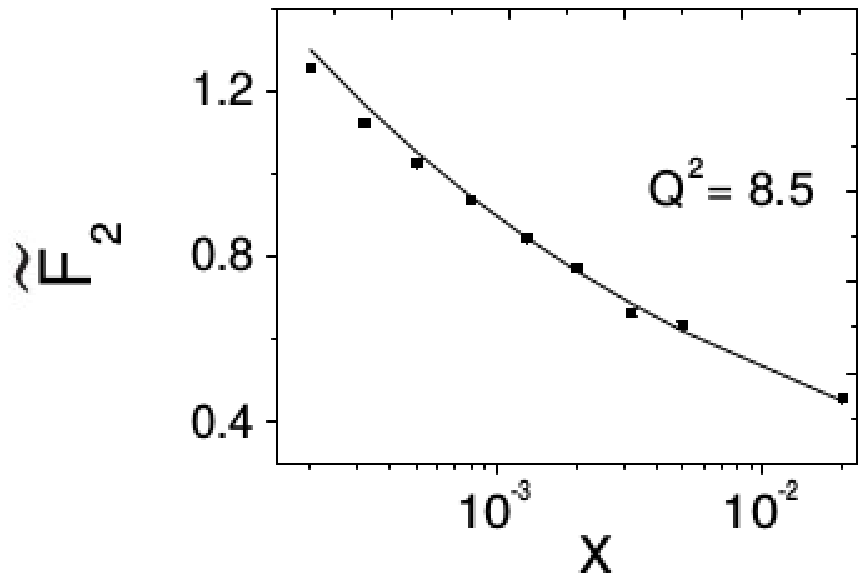}}\quad
\subfloat[]{\includegraphics[width=.29\textwidth]{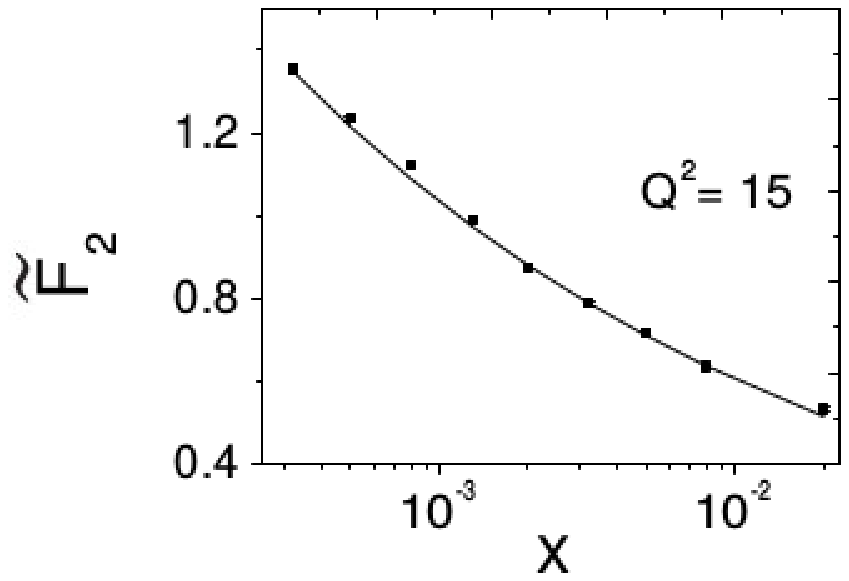}}\quad
 \subfloat[]{\includegraphics[width=.3\textwidth]{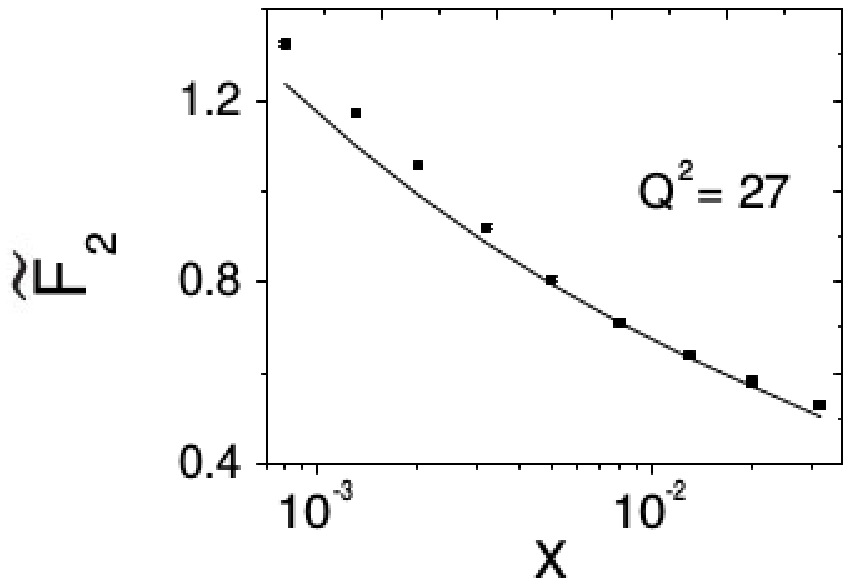}}\quad
 \subfloat[]{\includegraphics[width=.27\textwidth]{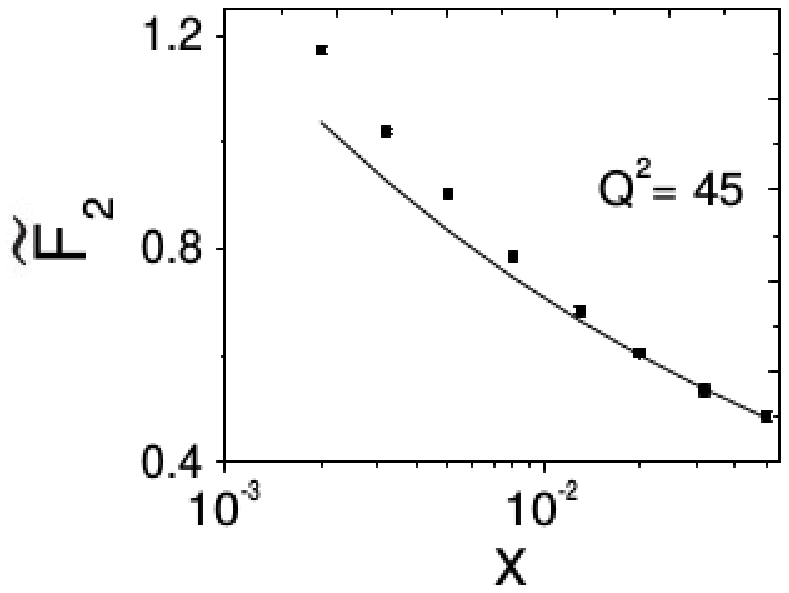}}\quad
  \subfloat[]{\includegraphics[width=.3\textwidth]{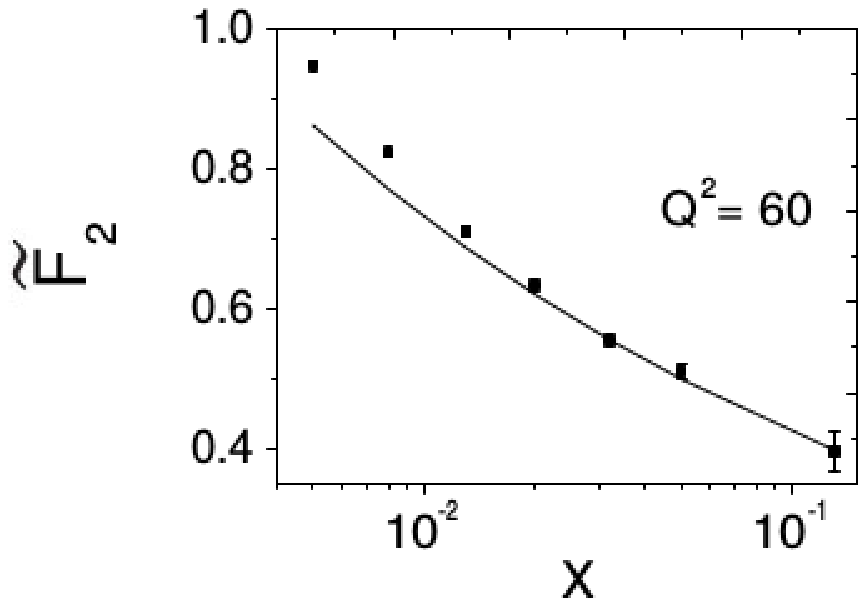}}\quad
\subfloat[]{\includegraphics[width=.31\textwidth]{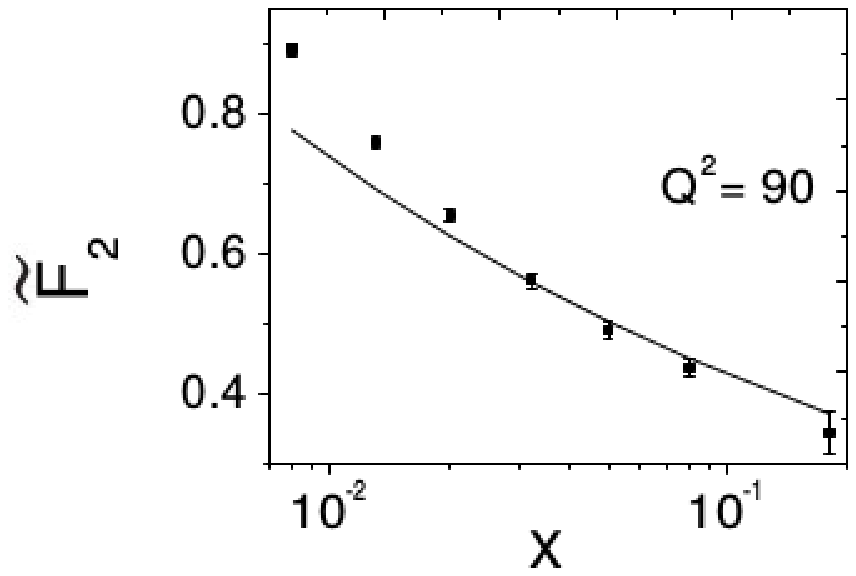}}\quad
\subfloat[]{\includegraphics[width=.3\textwidth]{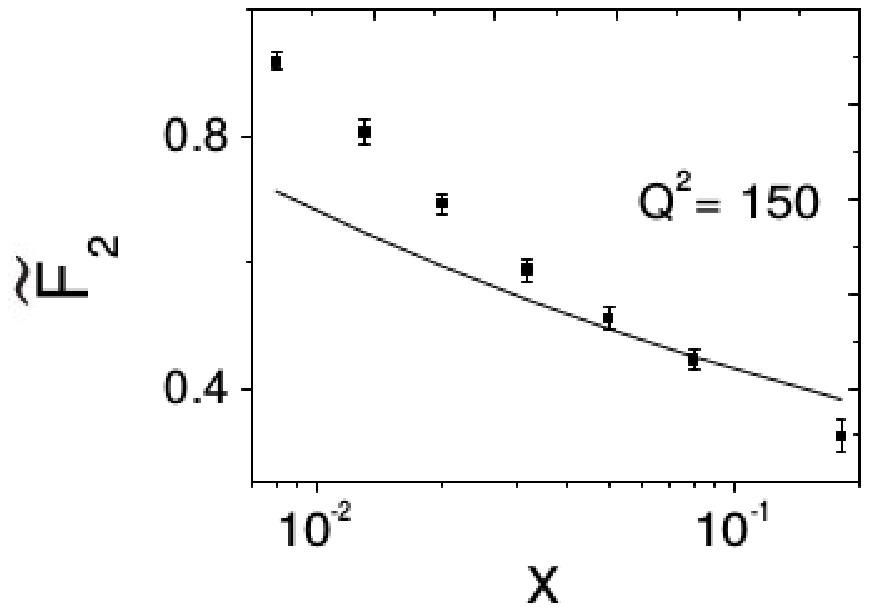}}\quad
 \subfloat[]{\includegraphics[width=.27\textwidth]{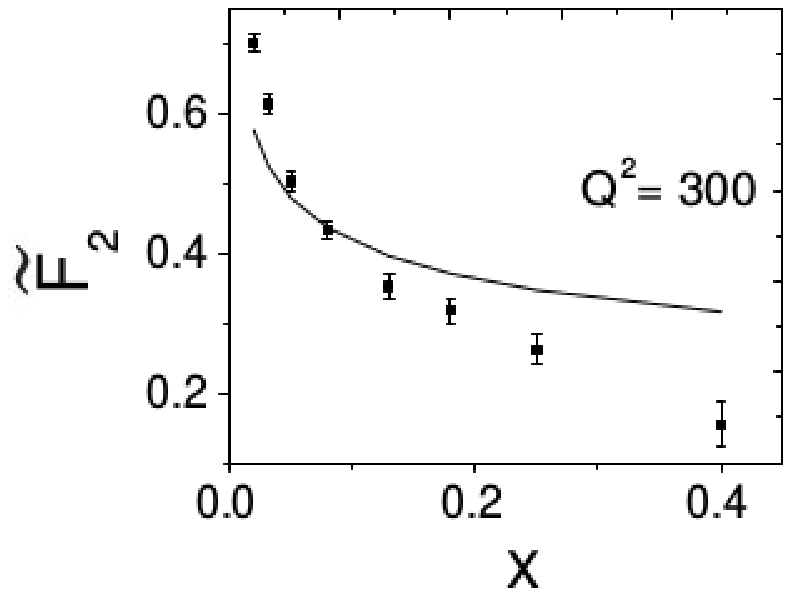}}\quad
 \subfloat[]{\includegraphics[width=.29\textwidth]{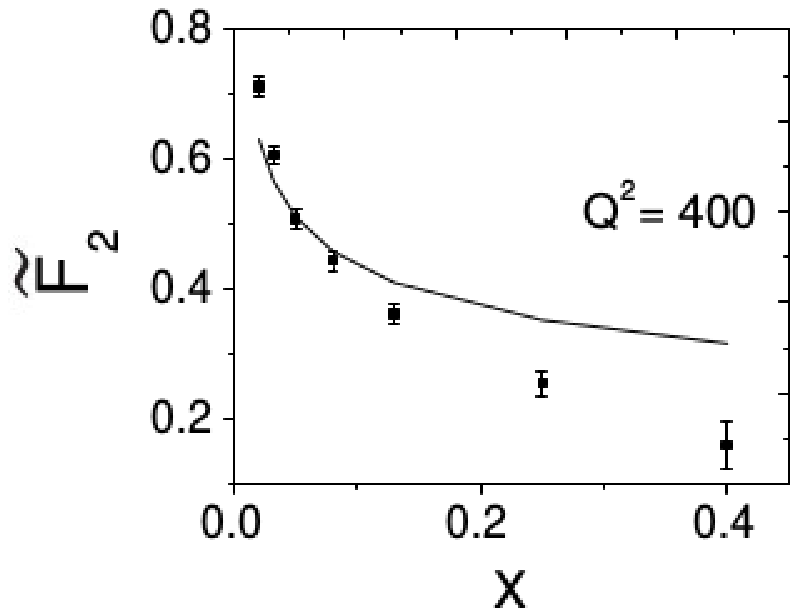}}\quad
\subfloat[]{\includegraphics[width=.29\textwidth]{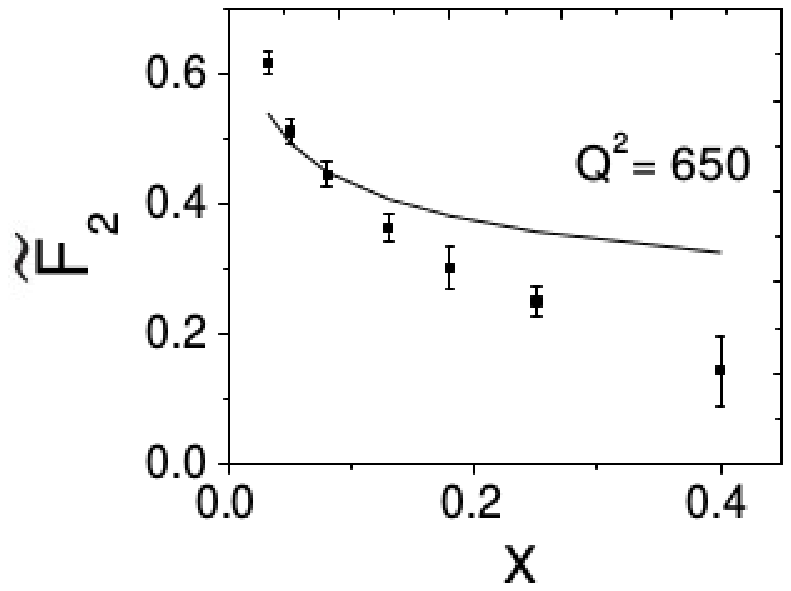}}\quad
\subfloat[]{\includegraphics[width=.31\textwidth]{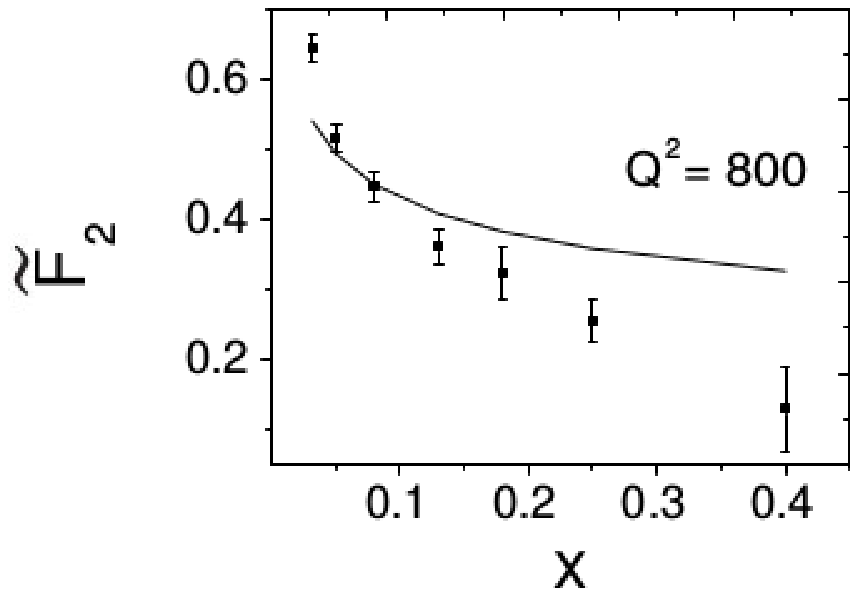}}\quad
 \caption{{\footnotesize comparison of the structure function $\tilde{F}_2$(Model 4) as a function of $x$ in bins of $Q^2$ with measured data of $F_2$ from HERAPDF1.0\cite{HERA}}}
 \label{F5}
\end{figure*}

\begin{figure*}[!tbp]
 \captionsetup[subfigure]{labelformat=empty}
\centering
  \subfloat[]{\includegraphics[width=.29\textwidth]{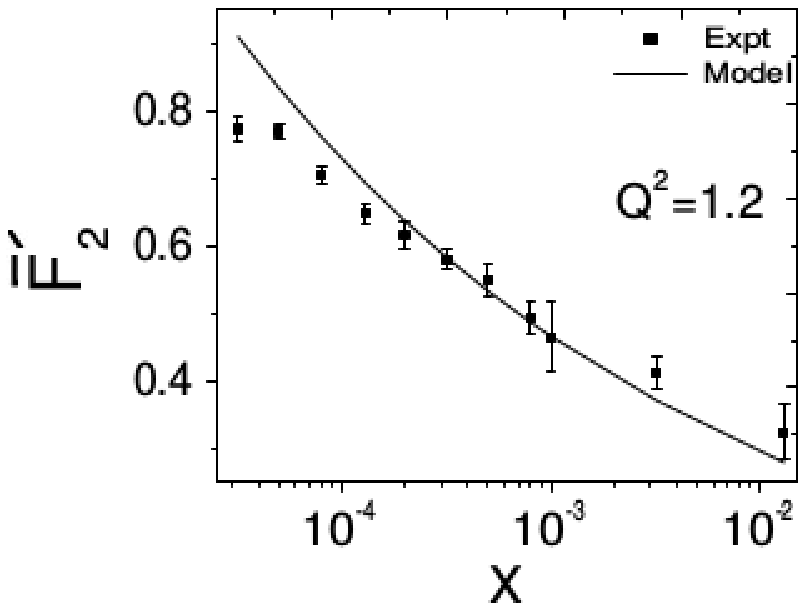}}\quad
 \subfloat[]{\includegraphics[width=.3\textwidth]{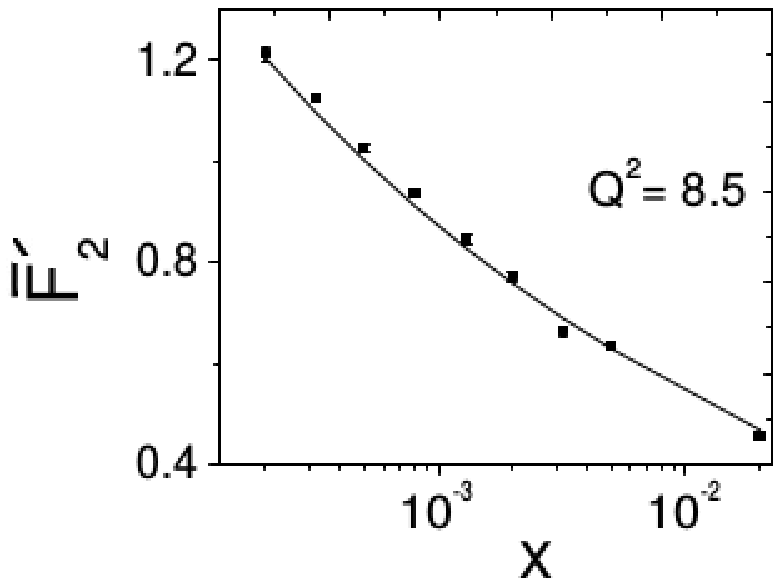}}\quad
\subfloat[]{\includegraphics[width=.3\textwidth]{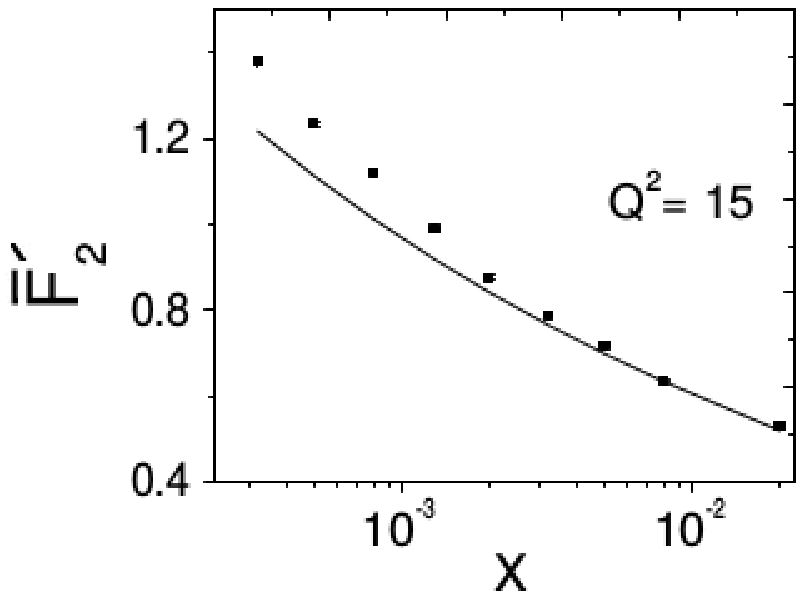}}\quad
 \subfloat[]{\includegraphics[width=.31\textwidth]{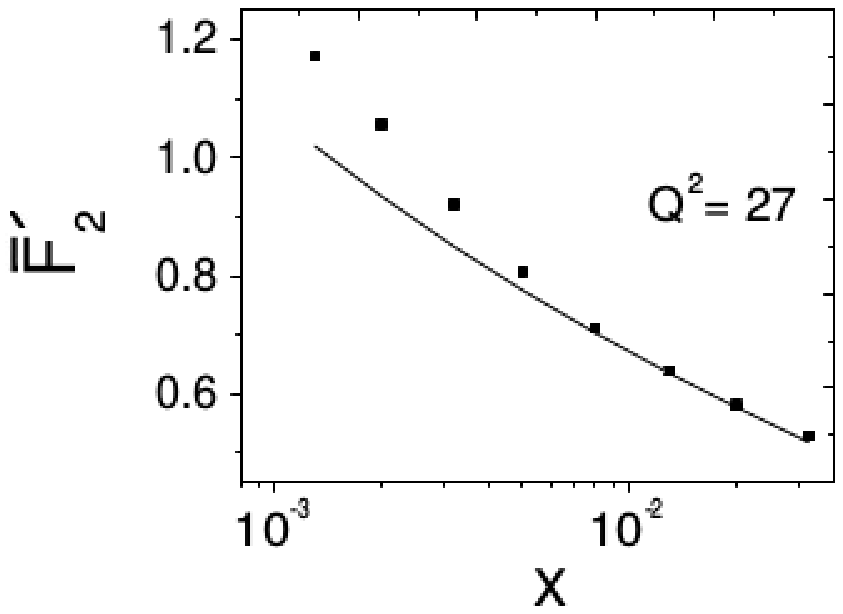}}\quad
  \subfloat[]{\includegraphics[width=.3\textwidth]{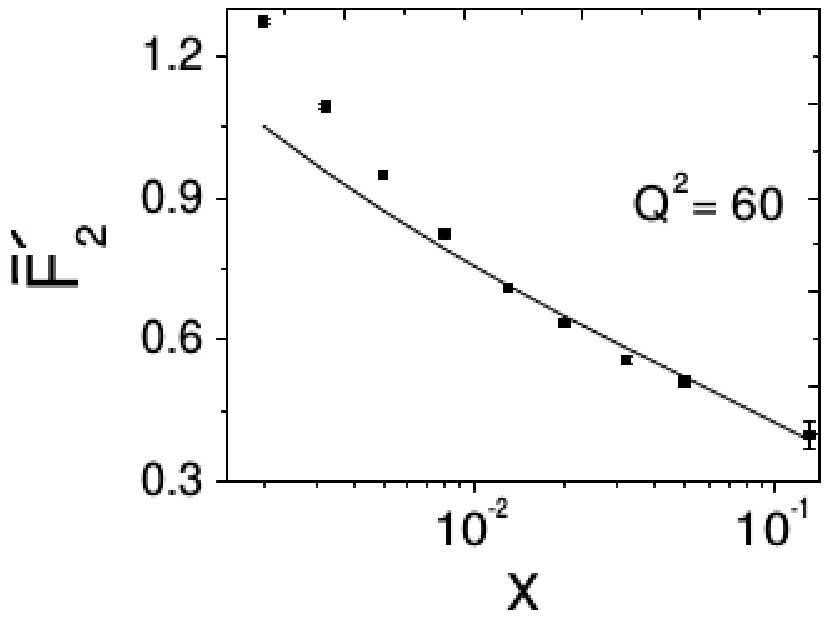}}\quad
\subfloat[]{\includegraphics[width=.3\textwidth]{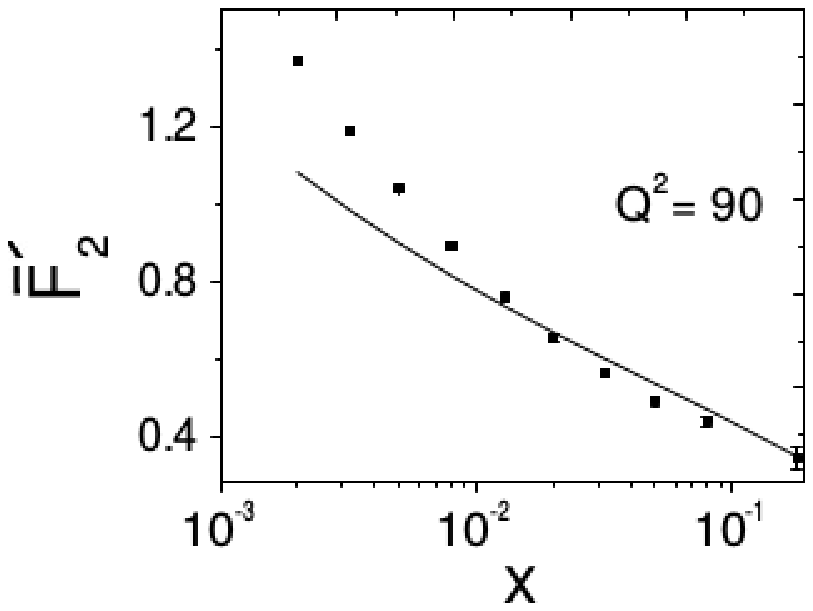}}\quad
\subfloat[]{\includegraphics[width=.3\textwidth]{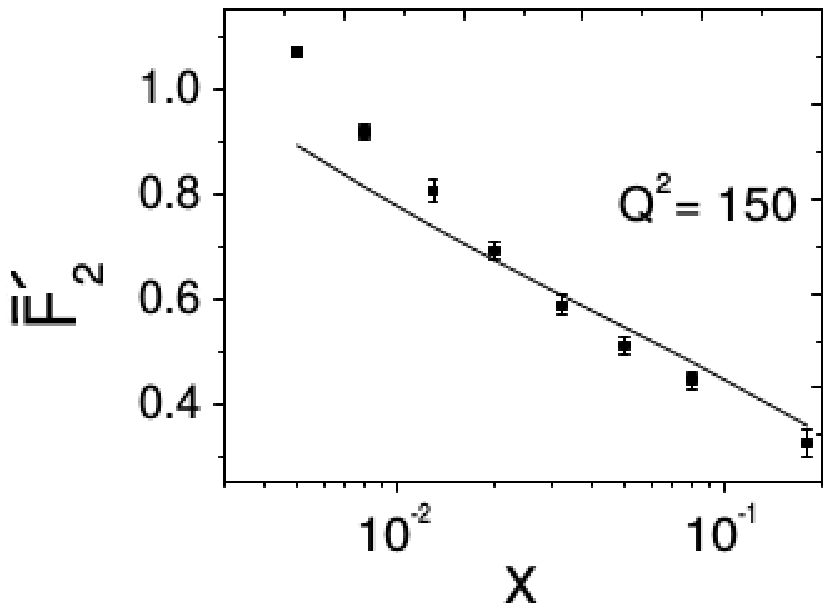}}\quad
 \subfloat[]{\includegraphics[width=.3\textwidth]{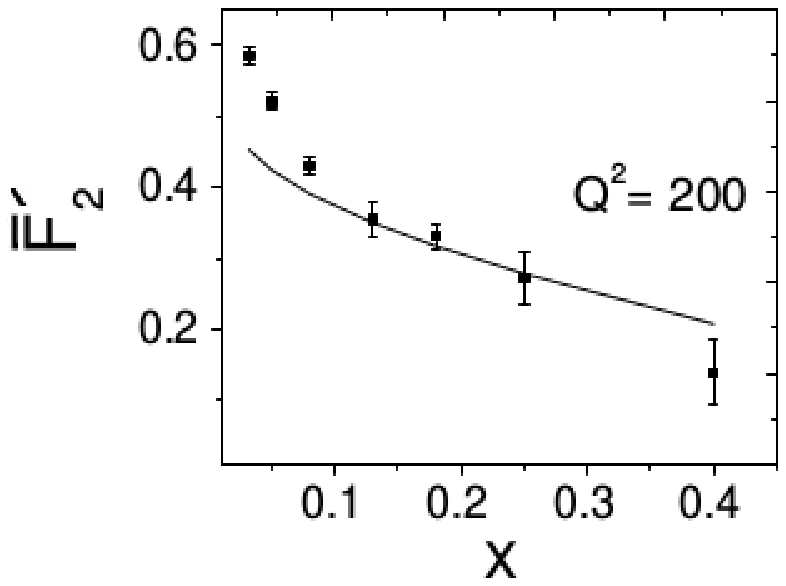}}\quad
 \subfloat[]{\includegraphics[width=.3\textwidth]{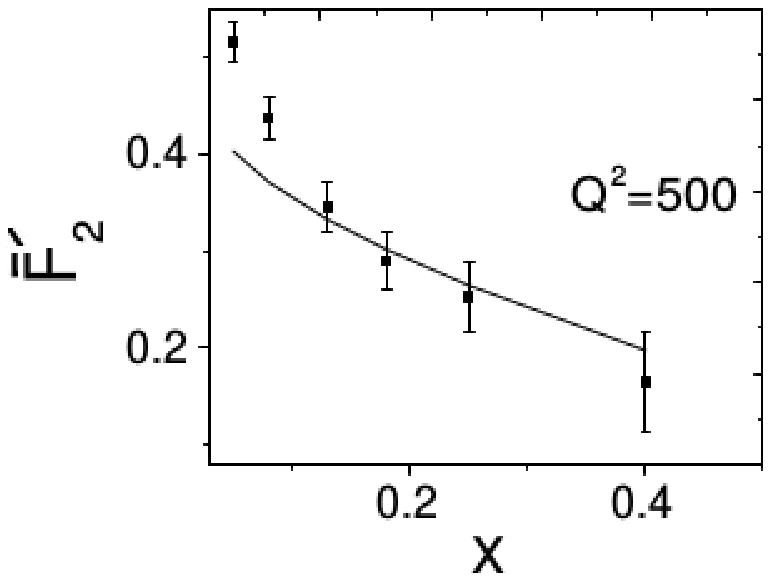}}\quad
\subfloat[]{\includegraphics[width=.3\textwidth]{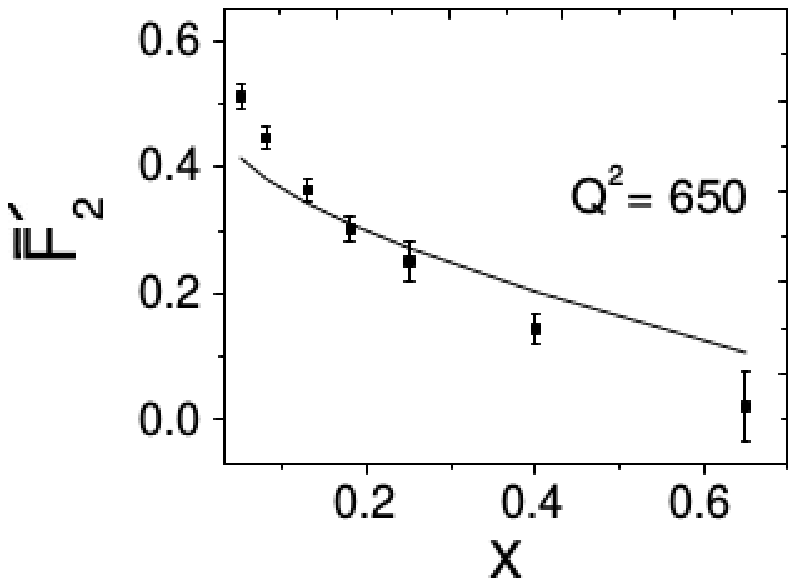}}\quad
\subfloat[]{\includegraphics[width=.3\textwidth]{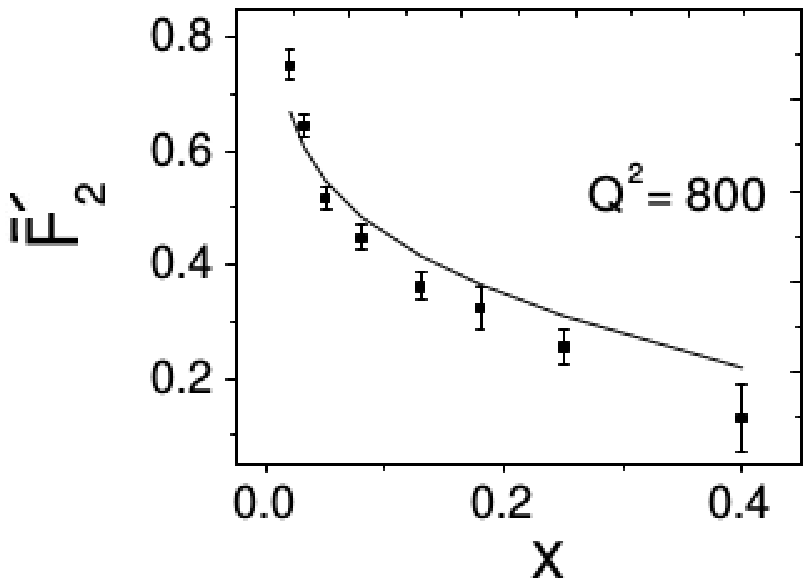}}\quad
 \subfloat[]{\includegraphics[width=.29\textwidth]{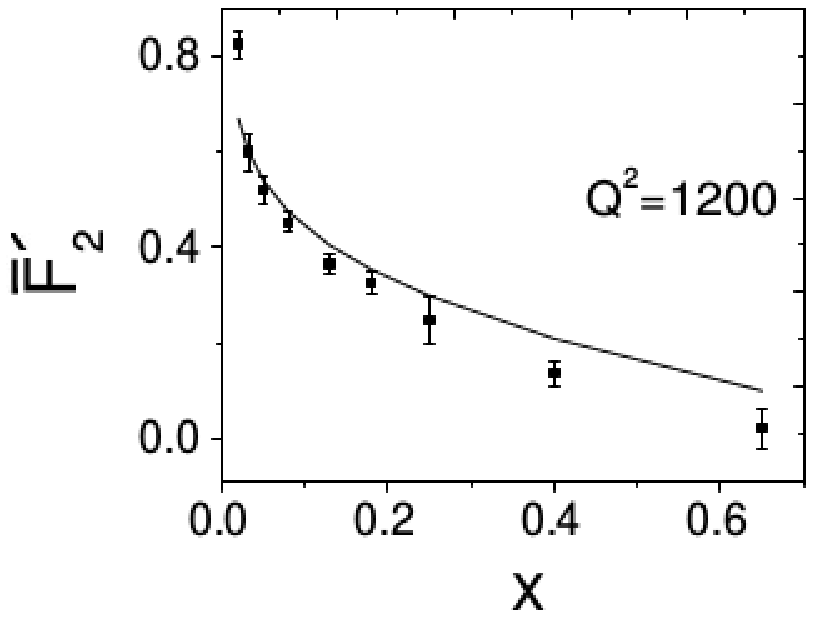}}\quad
 \caption{{\footnotesize comparison of structure function $\bar{F}'_2$(Model 6) as a function of $x$ in bins of $Q^2$ with measured data of $F_2$ from HERAPDF1.0\cite{HERA}}}
 \label{F6}
\end{figure*}

In Fig \ref{F5} and \ref{F6}, we plot $\tilde{F_2}$ and $\bar{F'_2}$ of Models 4 and 6 respectively as a function of \textit{x} for few representative values of $Q^2$ (Model 4: $Q^2= 1.2, 8.5, 15, 27, 45, 60, 90, 150, 300, 400, 650, 800$ GeV$^2$ and Model 6: $Q^2= 1.2, 8.5, 15, 27, 60, 90, 150, 200, 500, 650, 800, 1200$ GeV$^2$).
\begin{table}[!bp]
\caption{\label{T3}%
Results of the fit of Model 4}
\begin{ruledtabular}
\begin{tabular}{ccccccc}
\textrm{$\tilde{D}_0$}&
\textrm{$\tilde{D}_2$}&
\textrm{$\tilde{B}_1$}&
\textrm{$\tilde{B}_2$}&
\textrm{$\tilde{Q}_0^2$(GeV$^2$)}&
\textrm{$\chi^2$}&
\textrm{$\chi^2$/ndf} \\
\colrule
0.294\tiny${\pm 0.009}$ & 1.237\tiny${\pm 0.01}$ & 0.438\tiny${\pm 0.004}$ & 0.687\tiny${\pm 0.02}$ & 0.046\tiny${\pm 0.0004}$ & 170.616 & 0.60 \\
\end{tabular}
\end{ruledtabular}
\end{table}
\\
\begin{table}[!tbp]
\caption{\label{T4}%
Results of the fit of Model 6}
\begin{ruledtabular}
\begin{tabular}{ccccccc}
\textrm{$\bar{D}'_0$}&
\textrm{$\bar{D}'_2$}&
\textrm{$\bar{B}'_1$}&
\textrm{$\bar{B}'_2$}&
\textrm{$\bar{Q}_0'^2$(GeV$^2$)}&
\textrm{$\chi^2$}&
\textrm{$\chi^2$/ndf} \\
\colrule
0.335\tiny${\pm 0.003}$ & 1.194\tiny${\pm 0.0009}$ & 0.519\tiny${\pm 0.006}$ & 0.082\tiny${\pm 0.001}$ & 0.056\tiny${\pm 0.001}$ & 74.542 & 0.24\\
\end{tabular}
\end{ruledtabular}
\end{table}

\subsubsection{\textbf{Graphical representation of TMDPDFs of Model 4 and 6:}}
As in section \ref{3b}, we use the Eqn(20) as illustrated example and for $n_f=4$ the form of TMDs of Models 4 and 6 are:
\begin{equation}
\label{E22}
{\text M\text o\text d\text e\text l \ \text 4: } \quad  \tilde{f}_i(x,k_t^2)=\frac{e^{\tilde{D}_0^i}}{M^2}\left( \frac{1}{x} \right)^{\tilde{D}_2}\left( \frac{\tilde{B}_1}{\left( 1+\frac{k_t^2}{\tilde{k}_0^2} \right)} \right) \left( 1+ \frac{\tilde{B}_2}{\tilde{B}_1} \frac{1}{\left( 1+ \frac{k_t^2}{\tilde{k}_0^2} \right) }  \right)
\end{equation}
with $e^{{\tilde{D}_0}^u}=1.097=e^{{\tilde{D}_0}^d}$ and $e^{{\tilde{D}_0}^s}=0.274=e^{{\tilde{D}_0}^b}$.
\begin{equation}
\label{E22}
{\text M\text o\text d\text e\text l \ \text 6: } \quad   \bar{f}_i'(x,k_t^2)=\frac{e^{\bar{D}_0'^i}}{M^2}\left( \frac{1}{x} \right)^{\bar{D}'_2}\left(1-x \right)^{\bar{D}'_2}\left( \frac{\bar{B}'_1}{\left( 1+\frac{k_t^2}{\bar{k}_0'^2} \right)} \right) \left( 1+ \frac{\bar{B}'_2}{\bar{B}'_1} \frac{1}{\left( 1+ \frac{k_t^2}{\bar{k}_0'^2} \right) }  \right)
\end{equation}
with $e^{\bar{D}_0'^u}=1.14=e^{\bar{D}_0'^d}$ and $e^{\bar{D}_0'^s}=0.285=e^{\bar{D}_0'^b}$.

\begin{figure}[!bp]
\subfloat[]{%
  \includegraphics[width=0.49\linewidth]{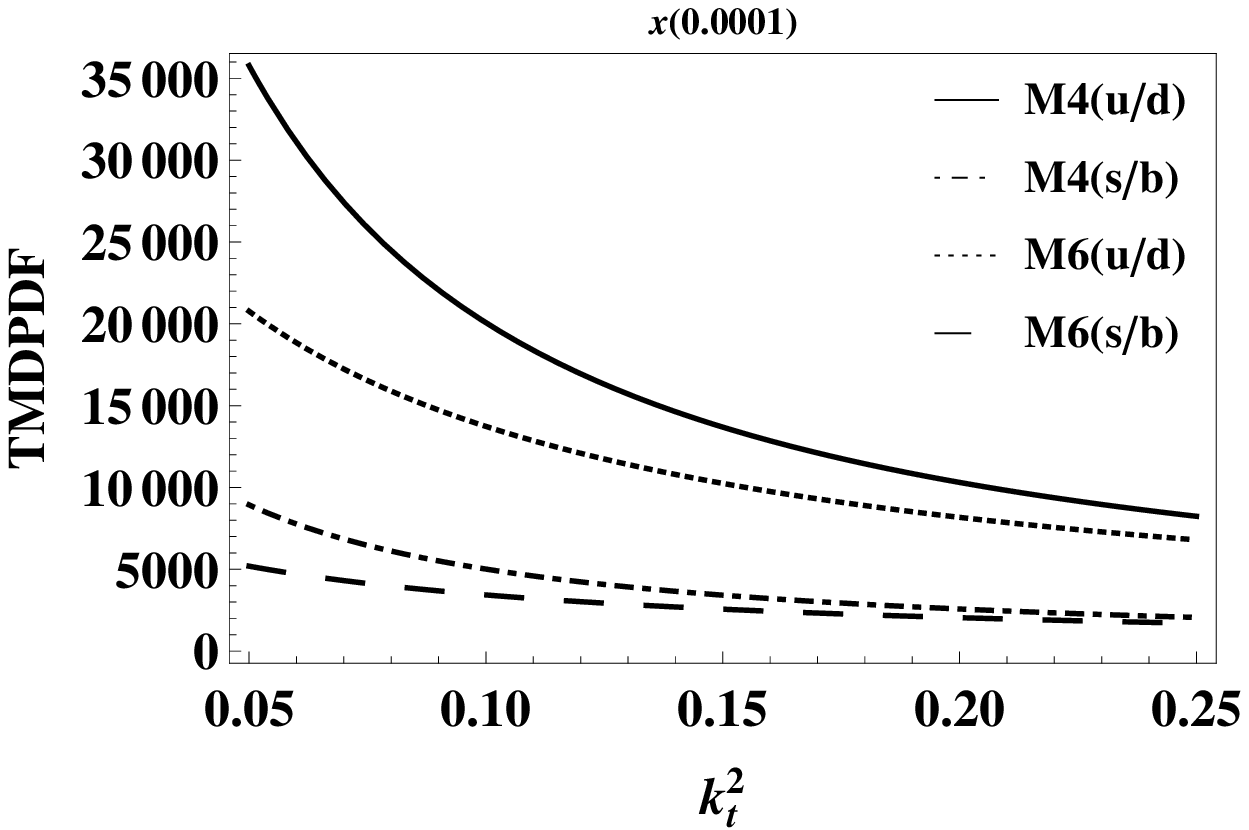}%
}\vspace{1ex}
\subfloat[]{%
  \includegraphics[width=0.46\linewidth]{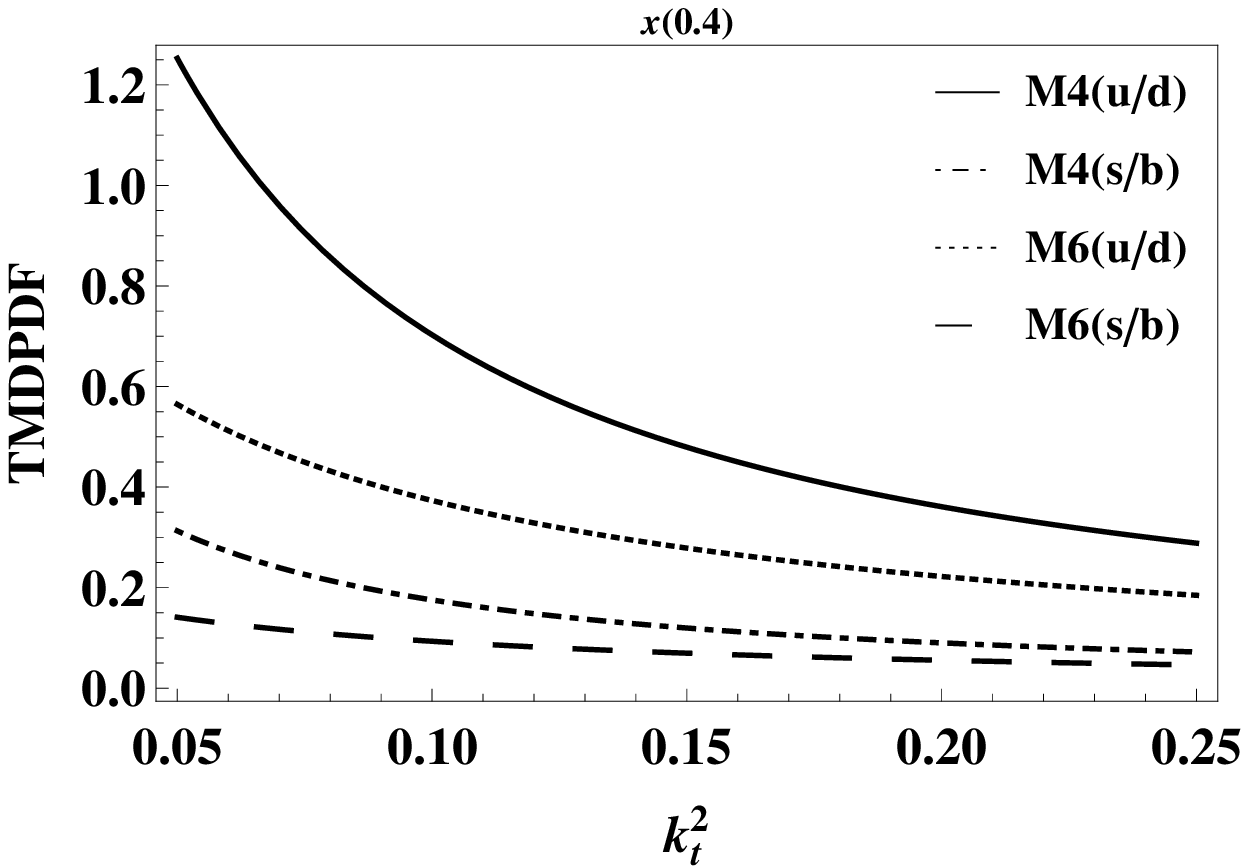}%
}\vspace{1ex}
\caption{TMDPDF vs $k_t^2$ for two representative values of (a) $x=10^{-4}$ and (b) $x=0.4$ for Models 4 and 6. Here, M4(u/d) (black line) and M6(u/d) (black dotted) represents the TMDPDF for u and d quarks for Models 4 and 6 respectively. Similarly, M4(s/b) (black dot-dashed) and M6(s/b) (black dashed) represents the TMDPDF for s and b quarks for Models 4 and 6 respectively.}
\label{F7}
\end{figure}

\begin{figure}[!tbp]
\subfloat[]{%
  \includegraphics[width=0.49\linewidth]{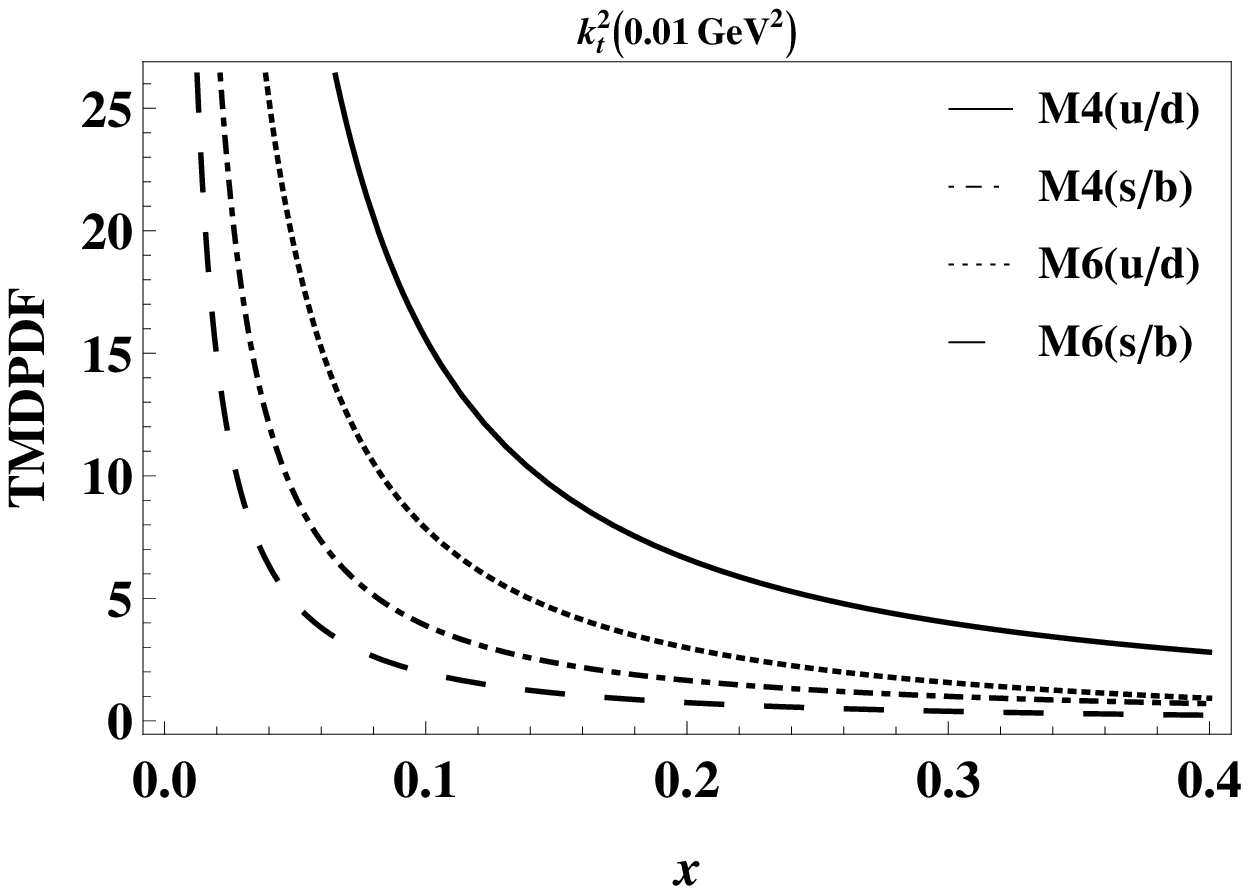}%
}\vspace{1ex}
\subfloat[]{%
  \includegraphics[width=0.48\linewidth]{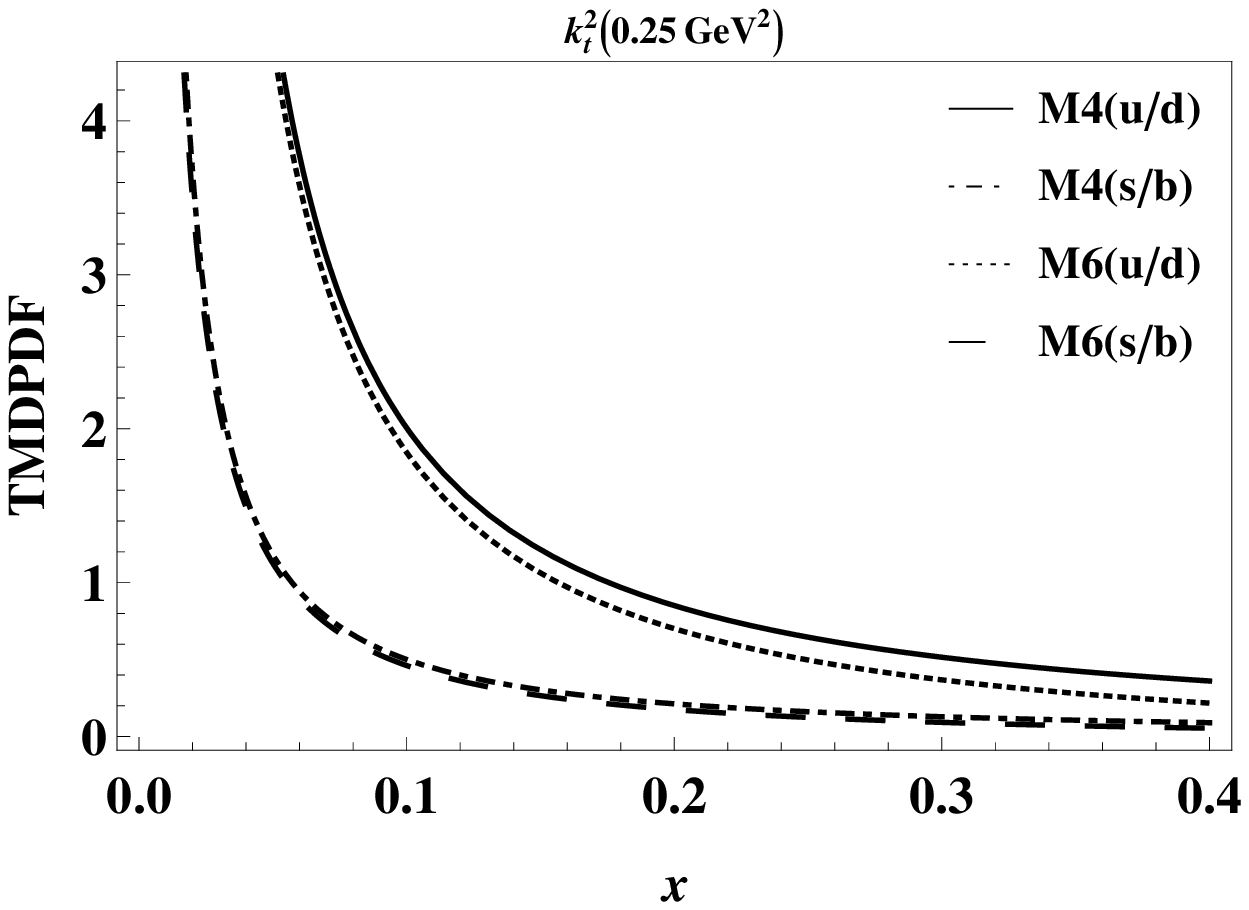}%
}\vspace{1ex}

\caption{TMDPDF vs $x$ for two representative values of (a) $k_t^2=0.01$ GeV$^2$ and (b) $k_t^2=0.25$ GeV$^2$ for Models 4 and 6. Here, M4(u/d) (black line) and M6(u/d) (black dotted) represents the TMDPDF for u and d quarks for Models 4 and 6 respectively. Similarly, M4(s/b) (black dot-dashed) and M6(s/b) (black dashed) represents the TMDPDF for s and b quarks for Models 4 and 6 respectively.}
\label{F8}
\end{figure}

Graphical representation of TMDPDFs of Model 4 and 6 are given in Fig \ref{F7} and \ref{F8}. It shows both the form of TMDs have got desired $k_t^2$ fall without the burden of singularities as expected. 

\begin{figure}[!tbp]
\subfloat[]{%
  \includegraphics[width=0.50\linewidth]{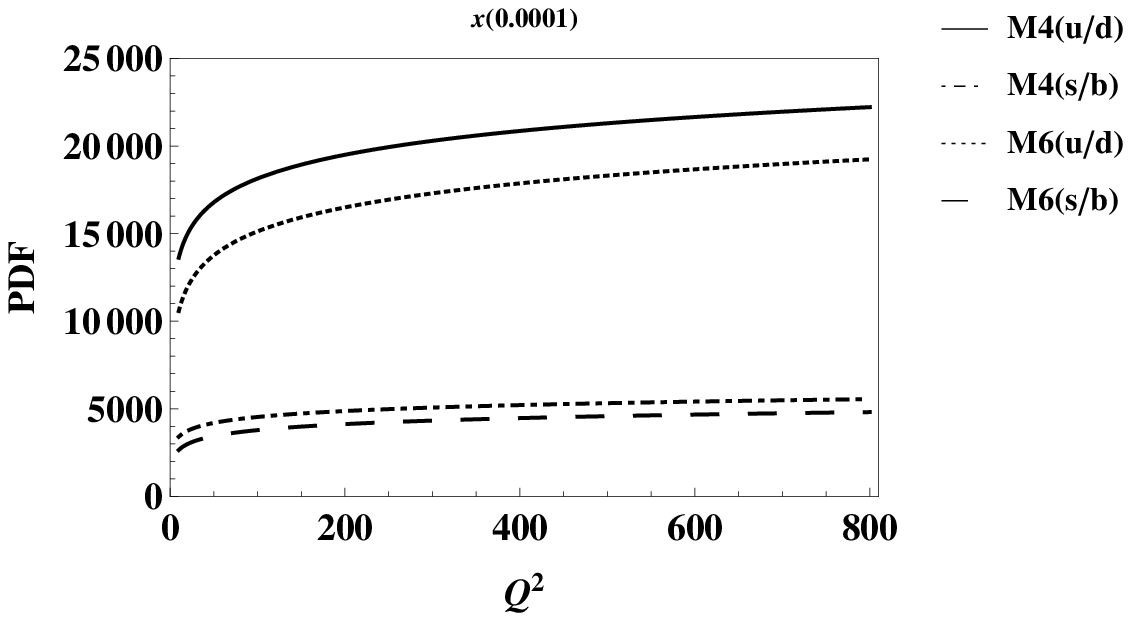}%
}\vspace{1ex}
\subfloat[]{%
  \includegraphics[width=0.47\linewidth]{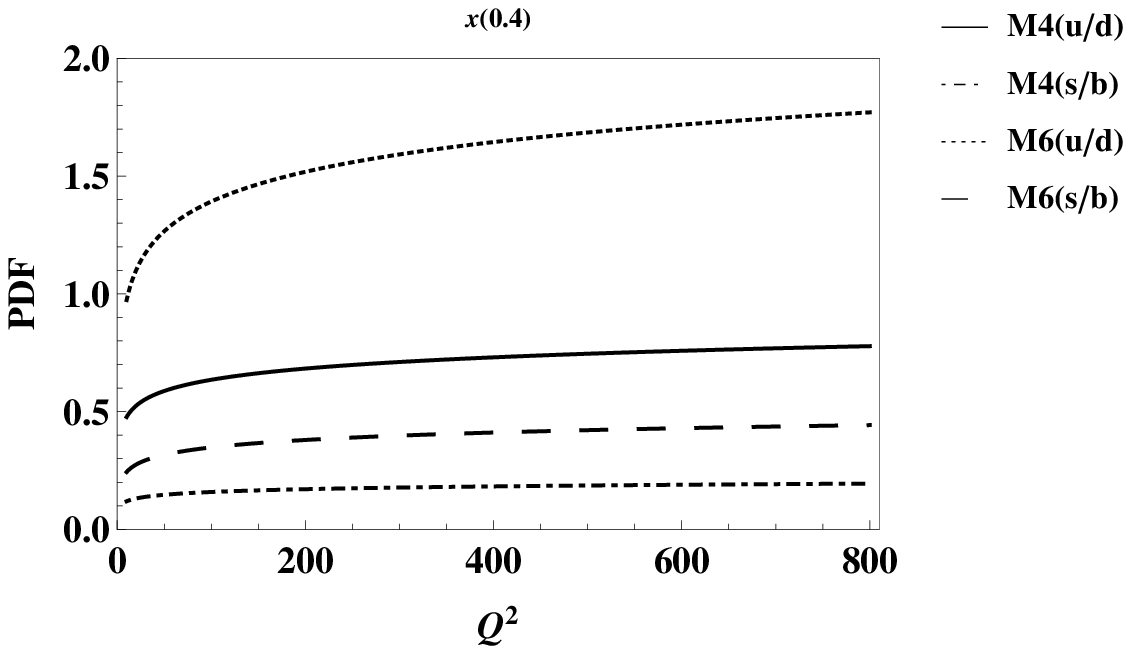}%
}\vspace{1ex}
\caption{PDF vs $Q^2$ for two representative values of (a) $x=10^{-4}$ and (b) $x=0.4$ for Models 4 and 6. Here, M4(u/d) (black line) and M6(u/d) (black dotted) represents the PDF for u and d quarks for Models 4 and 6 respectively. Similarly, M4(s/b) (black dot-dashed) and M6(s/b) (black dashed) represents the PDF for s and b quarks for Models 4 and 6 respectively.}
\label{F9}
\end{figure}

\begin{figure}[!tbp]
\subfloat[]{%
  \includegraphics[width=0.48\linewidth]{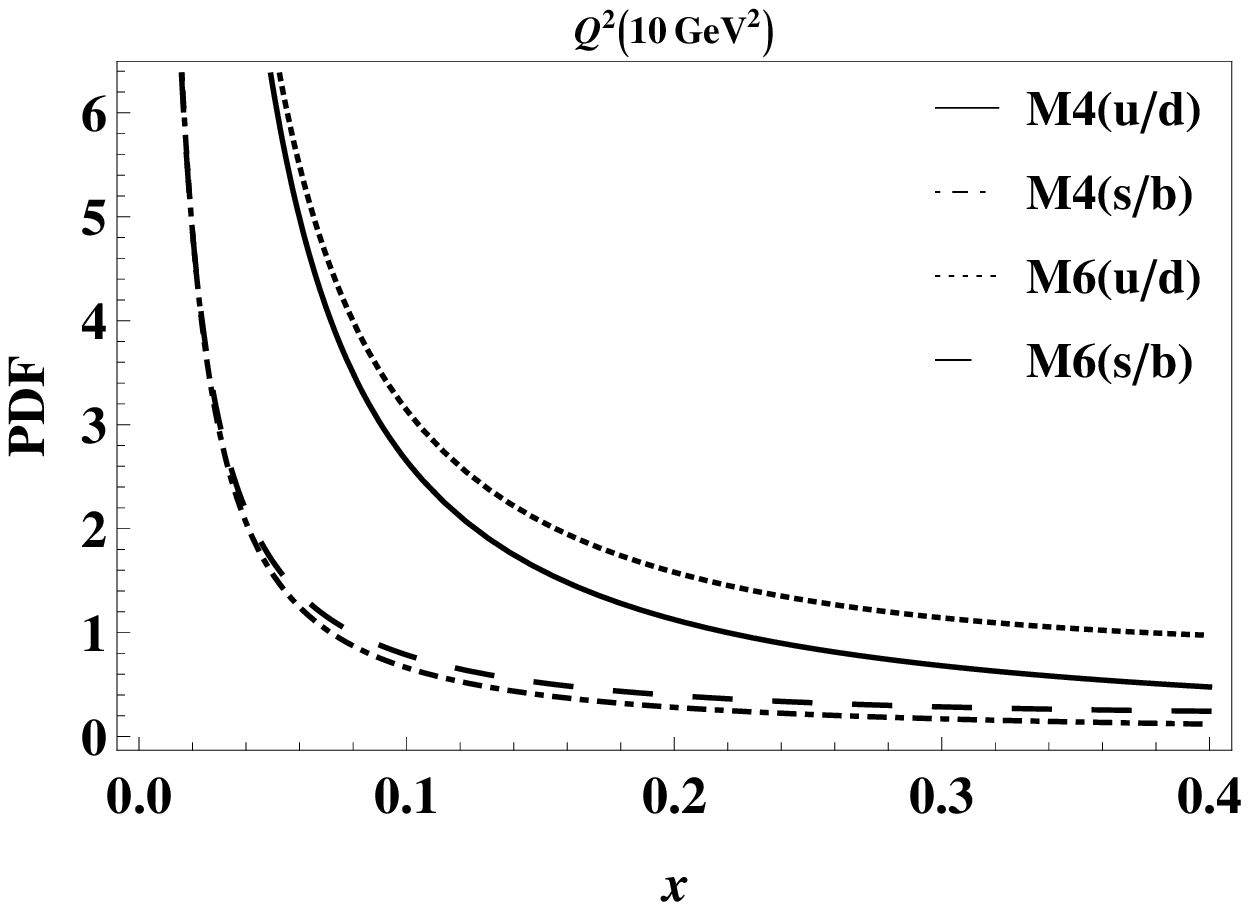}%
}\vspace{1ex}
\subfloat[]{%
  \includegraphics[width=0.49\linewidth]{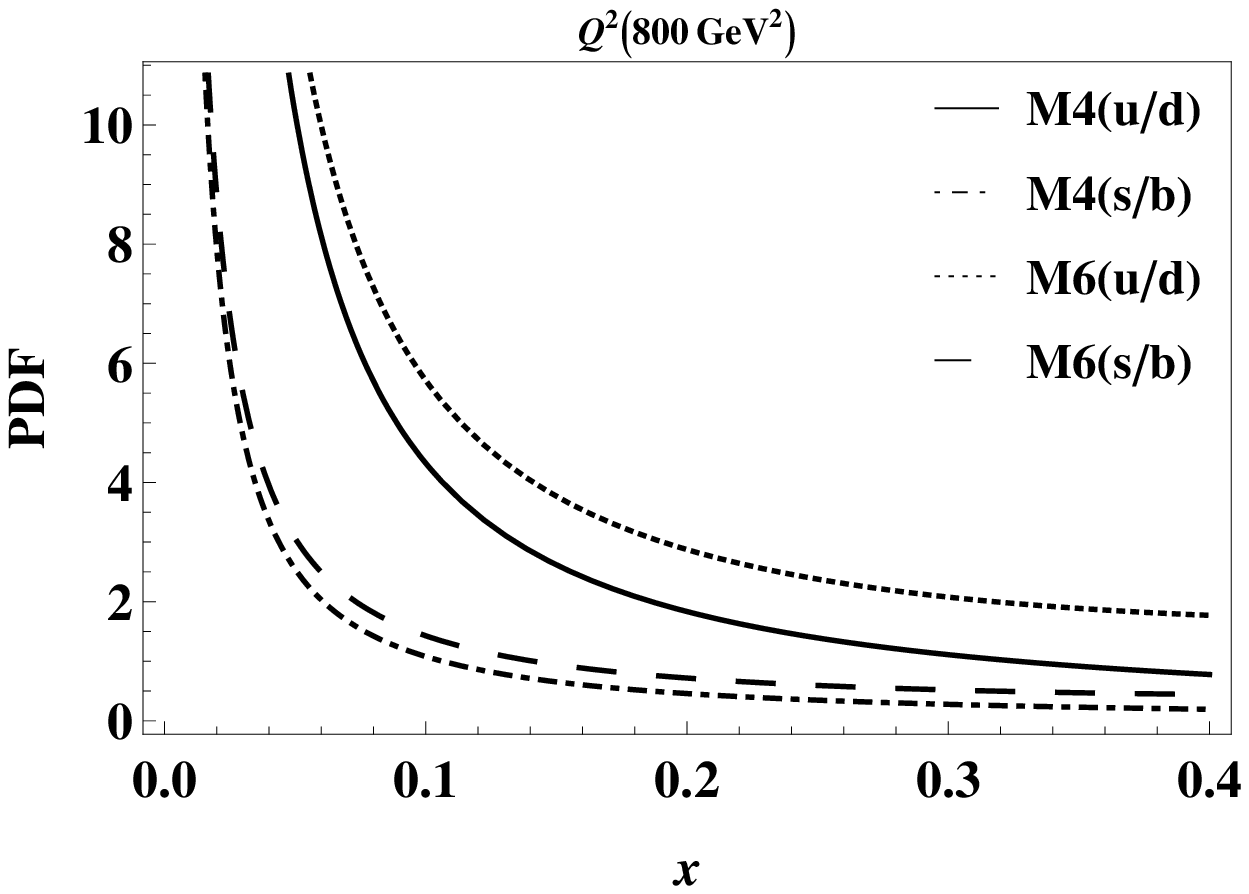}%
}\vspace{1ex}

\caption{PDF vs $x$ for two representative values of (a) $Q^2=10$ GeV$^2$ and (b) $Q^2=800$ GeV$^2$ for Models 4 and 6. Here, M4(u/d) (black line) and M6(u/d) (black dotted) represents the PDF for u and d quarks for Models 4 and 6 respectively. Similarly, M4(s/b) (black dot-dashed) and M6(s/b) (black dashed) represents the PDF for s and b quarks for Models 4 and 6 respectively.}
\label{F10}
\end{figure}

\subsubsection{\textbf{Graphical representation of PDFs of Models 4 and 6:}}
The given form of PDFs for Models 4 and 6 are:
\begin{eqnarray}
{\text M\text o\text d\text e\text l \ \text 4: } \quad \tilde{q}_i(x,Q^2) &=& \frac{e^{\tilde{D}_0^i} \tilde{Q}_0^2}{M^2}\left( \frac{1}{x} \right)^{\tilde{D}_2} \tilde{B}_1 \left[ \log \left( 1+\frac{Q^2}{\tilde{Q}_0^2} \right) -\frac{\tilde{B}_2}{\tilde{B}_1}\left(\frac{1}{\left( 1+\frac{Q^2}{\tilde{Q}_0^2}\right) }-1 \right)  \right] \\
{\text M\text o\text d\text e\text l \ \text 6: } \quad  \bar{q}'_i(x,Q^2) &=& \frac{e^{\bar{D}_0'^i} \bar{Q}_0'^2}{M^2}\left( \frac{1}{x} \right)^{\bar{D}'_2} \left(1-x \right)^{\bar{D}'_2} \bar{B}'_1 \left[ \log \left( 1+\frac{Q^2}{\bar{Q}_0'^2} \right) -\frac{\bar{B}'_2}{\bar{B}'_1}\left(\frac{1}{\left( 1+\frac{Q^2}{\bar{Q}_0'^2}\right) }-1 \right)  \right] \nonumber
\\
\end{eqnarray}
Graphical representation of PDFs of Model 4 and 6 are shown in Fig \ref{F9} and \ref{F10}. As expected both the models have proper $\log Q^2$ rise. The rise of the structure function at small \textit{x} is also compatible with Regge based models \cite{reg,m,boo,coo,yu}.
\section{Summary}
\label{E}
In this paper, we have made a comparative study of the two models of proton structure function based on self-similarity. The former model (Model 1) has got its singularity at $x_0\sim 0.019$ outside its phenomenological range of validity : \textit{x} : $6.2\times10^{-7}\leq x\leq 10^{-2}$ . The later one (Model 2) is completely free from singularity in the entire \textit{x}-range $0\leq x \leq 1$, but has a very restrictive phenomenological range of validity : $Q^2$ : $0.85\leq Q^2 \leq$ 10 GeV$^2$. At phenomenological level, the former one is thus better than later. The corresponding TMDs are then studied and the main difference is observed : the singularity free version of the model (Model 2) results in a TMD which does not have the expected qualitative feature, while the other one (Model 1) has. In order to remove the anomalies, we have generalized the definition of a defining magnification factor in TMD such that it has expected qualitative feature (Model 3). In a specific case, where the defining parameters of the generalized TMD satisfy certain specific conditions among them, then the resulting structure function (Model 4) has also logarithmic rise in $Q^2$ as expected in any QCD compatible model, instead of power laws of previous models (Model 1, 2, and 3) without addition of any new parameters. The model has now got larger phenomenological range of validity in $Q^2$ than the earlier ones.

Assuming that the notion of self-similarity can be smoothly extrapolated into larger \textit{x}, we have also obtained a model at large and small \textit{x} (Model 5) for TMDPDF/PDF and structure function. As in previous case at small \textit{x} (Model4), under specific condition amongst its   model parameters, $\log Q^2$ rise in the resulting structure function (Model 6) emerges. The extrapolated model has also been tested with combined HERA data \cite{HERA} and wider phenomenological range of $x$ and $Q^2$ has been obtained as expected.

Let us end this section with the theoretical limitation of the present work. As noted in the introduction, self-similarity is not a general property of QCD and is not yet established either, theoretically or experimentally. In this work, we have merely used the notion of self-similarity in parametrize TMDs and PDFs as a generalization of the method suggested in Ref\cite{Last} and have shown that under specific conditions among the defining parameters, logarithmic rise in $Q^2$ of structure function is achievable even in such an approach, compatible with QCD expectation and has wider phenomenological ($x-Q^2$) range of validity. It presumably implies that while self-similarity has not yet been proven to be a general feature of QCD, under specific conditions, experimental data can be interpreted with this notion as has been shown in the present paper. However, to prove it from the first principle is beyond the scope of the present work.

\section*{Acknowledgment}
We thank Dr. Kushal Kalita for helpful discussions, Dr. Rupjyoti Gogoi of Tezpur University for collaboration at the initial stage of the work and Dr. Akbari Jahan for useful comments on self-similarity. One of the authors (BS) acknowledges the UGC-RFSMS for financial support. Final part of this work was completed when one of us (DKC) visited the Rudolf Peirels Center of Theoretical Physics, University of Oxford. He thanks Professor Subir Sarkar and Professor Amanda Cooper-Sarkar for useful discussion.

\end{document}